\documentclass[aps,pra,twocolumn,superscriptaddress,floatfix,amsmath,amssymb,notitlepage]{revtex4-1}

\usepackage{graphicx}
\usepackage{enumitem}

\usepackage[caption=false]{subfig}
\usepackage{amsmath}	
\usepackage{amssymb}    
\usepackage[pdftex]{hyperref}

\newcommand{\be}{\begin{eqnarray}}
 \newcommand{\ee}{\end{eqnarray}}

 \newcommand{\bra}[1]{\left\langle #1 \right|}
 \newcommand{\ket}[1]{\left| #1 \right\rangle}

\begin{document}



\title{Time-domain quantum interference in graphene}

\author{Fran\c{c}ois Fillion-Gourdeau}
\email{francois.fillion@emt.inrs.ca}
\affiliation{Universit\'{e} du Qu\'{e}bec, INRS-\'{E}nergie, Mat\'{e}riaux et T\'{e}l\'{e}communications, Varennes, Qu\'{e}bec, Canada J3X 1S2}
\affiliation{Institute for Quantum Computing, University of Waterloo, Waterloo, Ontario, Canada, N2L 3G1}

\author{Denis Gagnon}
\affiliation{Universit\'{e} du Qu\'{e}bec, INRS-\'{E}nergie, Mat\'{e}riaux et T\'{e}l\'{e}communications, Varennes, Qu\'{e}bec, Canada J3X 1S2}
\affiliation{Institute for Quantum Computing, University of Waterloo, Waterloo, Ontario, Canada, N2L 3G1}

\author{Catherine Lefebvre}
\affiliation{Universit\'{e} du Qu\'{e}bec, INRS-\'{E}nergie, Mat\'{e}riaux et T\'{e}l\'{e}communications, Varennes, Qu\'{e}bec, Canada J3X 1S2}
\affiliation{Institute for Quantum Computing, University of Waterloo, Waterloo, Ontario, Canada, N2L 3G1}

\author{Steve MacLean}
\email{steve.maclean@emt.inrs.ca}
\affiliation{Universit\'{e} du Qu\'{e}bec, INRS-\'{E}nergie, Mat\'{e}riaux et T\'{e}l\'{e}communications, Varennes, Qu\'{e}bec, Canada J3X 1S2}
\affiliation{Institute for Quantum Computing, University of Waterloo, Waterloo, Ontario, Canada, N2L 3G1}

\date{\today}

\begin{abstract}
The electron momentum density obtained from the Schwinger-like mechanism is evaluated for a graphene sample immersed in a homogeneous time-dependent electric field. Based on the analogy between graphene low-energy electrons and quantum electrodynamics (QED), numerical techniques borrowed from strong field QED are employed and compared to approximate analytical approaches. It is demonstrated that for some range of experimentally accessible parameters, the pair production proceeds by sequences of adiabatic evolutions followed by non-adiabatic Landau-Zener transitions, reminiscent of the Kibble-Zurek mechanism describing topological defect density in second order phase transitions. For some field configurations, this yields interference patterns in momentum space which are explained in terms of the adiabatic-impulse model and the Landau-Zener-St\"{u}ckelberg interferometry.
\end{abstract}


\maketitle


\section{Introduction}

Graphene, a two-dimensional arrangement of carbon atoms on a honeycomb lattice structure, is a relatively new material which exhibits spectacular electronic \cite{geim2007rise,RevModPhys.81.109}, optical \cite{Bonaccorso2010, Yao2013}, and mechanical properties \cite{Lee385}.
These properties make graphene promising for the development of electronic and optoelectronic devices such as ballistic transistors \cite{geim2007rise}, solar cells \cite{Bonaccorso2010}, and photodetectors \cite{Schall2014}.
In addition to practical applications, the contribution of graphene to fundamental physics has also been recognized.
Specifically, charge transport in this 2D material is of particular interest because it is analogous to quantum electrodynamics (QED) \cite{gusynin2007ac}: in the low-energy limit of the tight-binding model, electrons propagating on the honeycomb lattice can be described by an effective theory based on a massless 2D Dirac equation \cite{PhysRev.71.622,RevModPhys.81.109}. The latter is similar to the Dirac equation governing the relativistic quantum behavior of the electron, except for the fact that the speed of light $c$ is replaced by the Fermi velocity $v_{F} =  1.093\times 10^{6}$ m/s. Also, the interaction with the electromagnetic sector is slightly different: in graphene, two-dimensional quasi-particles interact with three-dimensional photons while in QED, both electrons and photons ``live'' in the same number of dimensions. The theory describing graphene is thus massless reduced quantum electrodynamics (RQED$_{3,2}$, where the index denotes the photon and fermion dimensions, respectively) \cite{PhysRevD.86.025005}. Owing to this analogy, graphene can be used as a QED simulator if the following minimal set of experimental conditions is fulfilled:
\begin{enumerate}[label=(\arabic*)]
\item intrinsic graphene,
\item ``perfect'' lattice structure and relatively large domains,
\item small thermal effects,
\item small phonon dispersion,
\item small coupling constant,
\item momentum of quasi-particles close to Dirac points.
\end{enumerate}
These conditions are now discussed.
(1) The Fermi energy has to be precisely at the Dirac point to simulate the QED vacuum. By definition, intrinsic graphene obeys this property but may be challenging to produce experimentally because a small external potential or doping will induce charge carriers in the conduction band \cite{Adam20112007}. Nevertheless, QED can still be simulated if the carrier density generated in this way is negligible compared to the process under consideration, as $\langle \tilde{n}_{\mathrm{carriers}} \rangle \ll \langle \tilde{n} \rangle$.  Henceforth, $\langle \tilde{n} \rangle$ characterizes the creation of electron-hole pairs by the Schwinger-like mechanism, which will be described in details in the next section.
(2) The presence of impurities and scattering on domain boundaries can change the behavior of quasi-particles, thereby affecting the transport properties of graphene  \cite{Xie2016}.
This implies using single crystal domains in the realization of a QED simulator with a size $\ell$ larger than the typical distance travelled by quasi-particles, i.e. $\ell \gg v_{F} t_{\mathrm{travel}}$, with $t_{\mathrm{travel}}$ the characteristic travelling time scale of quasi-particles.
Single crystal graphene with domain sizes as large as 20 $\mu$m can be fabricated using currently available technology \cite{Basov2014}.
(3) Intrinsic graphene at non-zero temperature $T$ has electrons in the conduction band due to thermal effects. The resulting electronic density, given by \cite{1.2776887}
\begin{eqnarray}
\langle \tilde{n}_{\mathrm{thermal}}\rangle  = \frac{\pi}{6}\left(\frac{k_{B}T}{\hbar v_{F}} \right)^{2} \ll \langle \tilde{n} \rangle,
\end{eqnarray} 
where $k_{B}$ is Boltzmann's constant, should be smaller than the one produced by the process under consideration ($\langle \tilde{n} \rangle$).  
(4) Phonons can interact with quasi-particles through a gauge-like coupling, introducing an additional scattering channel \cite{Xie2016}.
This scattering channel mainly results in a renormalization of the graphene Fermi velocity.
The impact of phonon coupling on quasi-particle lifetime can be mitigated by performing experiments at low temperature \cite{Park2007}. 
(5) Fermion interactions can be neglected at leading order like in QED when the coupling constant is small, i.e. when $g := \alpha c/\epsilon v_{F} \ll 1$, where $\alpha \approx 1/137$ is the fine-structure constant and $\epsilon$ is the dielectric constant of the substrate \cite{gusynin2007ac,PhysRevLett.113.105502,PhysRevB.89.235431}. 
From this power counting argument, suspended graphene ($g \simeq 2.3$) would not be suited to the experimental realization of a QED simulator. Instead, embedding graphene layers in a medium with a sufficiently high dielectric constant $\epsilon$ is required.
For instance, graphene deposited on SiO$_2$ \cite{doi:10.1021/jp5070215} yields a value of $g \simeq 0.9$, and substrates with higher dielectric constants are currently available \cite{Basov2014}. However, even for relatively large value of the coupling constant ($g \lesssim 2.3$), graphene may behave as a weakly coupled system where the Fermi velocity is renormalized \cite{PhysRevLett.113.105502}. 
(6) The Dirac points are positioned at the absolute momentum $|\mathbf{K}_{\pm}| = \frac{4\pi}{3\sqrt{3} a} \approx 3361$ eV, where $a \approx 1.42 \times 10^{-10}$~m is the distance between carbon atoms \cite{RevModPhys.83.407,RevModPhys.81.109}. Close to these points, the dispersion relation is linear and is given by
\begin{eqnarray}
E_{\mathbf{p}} = v_{F}|\mathbf{p}| + O(|\mathbf{p}|/|\mathbf{K}_{\pm}|),
\end{eqnarray}
where $\mathbf{p}$ is the relative momentum of quasi-particles measured with respect to $\mathbf{K}_{\pm}$ (here, the subscript $\pm$ refers to non-equivalent Dirac points, as described in more detail below). The dispersion relation holds when $|\mathbf{p}| \ll |\mathbf{K}_{\pm}|$. In this article, the maximum momentum of quasi-particles is estimated to be $|\mathbf{p}_{\rm max}| \approx 100$ eV, ensuring that quasi-particles have a linear dispersion relation. We note that a similar value for $|\mathbf{p}_{\rm max}|$ is found in Ref. \cite{RevModPhys.83.407}.

Although the above-mentioned conditions are stringent, they may be more easily achieved experimentally than certain requirements for the study of QED processes. For instance, the study of Schwinger's mechanism, whereby the vacuum decays into electron-positron pairs in the presence of a strong classical constant electromagnetic field, requires field strengths of $E_{0} \sim E_{S} := \frac{m^{2} c^{3}}{e\hbar} \approx 1.3 \times 10^{18}$ V/m (here, $m$ is the electron mass). Fields of this magnitude are unattainable with current laser technology: the highest field strengths attained are approximately given by $E_{\rm exp} \sim 10^{13}-10^{14}$ V/m \cite{RevModPhys.78.309}. The probability to create a pair in vacuum is given by \cite{PhysRev.82.664,Itzykson:1980rh}
\begin{eqnarray}
\label{eq:proba}
P_{S} \sim e^{- \pi \frac{E_{S}}{E_{0}}}.
\end{eqnarray} 
Therefore, there is an exponential suppression of the rate proportional to the mass gap $\Delta_{\mathrm{gap}}=mc^{2}$. In graphene, the quasi-particles are massless, reducing considerably the field strength required to produce electron-hole pairs \cite{PhysRevD.78.096009,PhysRevB.81.165431,PhysRevLett.102.106802,PhysRevD.87.125011,
PhysRevB.81.041416,PhysRevB.92.035401}. This and the fact that it is a QED simulator make graphene a good candidate to study Schwinger-like processes. As a matter of fact, several other QED-like phenomena have been investigated in graphene \cite{PhysRevLett.53.2449,katsnelson2007graphene,PhysRevLett.102.026807} because these processes are important to understand the conductivity and other properties of this material.

In this article, the process of electron-hole pair production in graphene is investigated using analytical and numerical methods in strong field RQED$_{3,2}$. Similar studies have been performed in the past \cite{PhysRevB.81.165431,PhysRevLett.102.106802,PhysRevD.87.125011,
PhysRevB.81.041416,PhysRevB.92.035401} but the phenomenon of quantum interference between each half-cycle of an oscillating field was generally overlooked as constant fields were generally considered (the so-called $T$-constant field \cite{PhysRevD.86.125022}). An oscillating field has been considered in p-n graphene junctions where it was demonstrated that quantum interferences are responsible for the current asymmetry \cite{PhysRevB.88.241112}. A similar applied field was investigated in Ref. \cite{akal2016low}, where a mass gap was considered. It was shown that a strong resonance behavior can be observed in the electron-hole pair momentum spectrum.
Pair production with graphene Landau levels (i.e. in the presence of a quantizing magnetic field) driven by circularly and linearly polarized fields has also been investigated \cite{Gagnon2016}.

In strong field QED, quantum interference is an important topic because it explains the peak and valley structure seen in numerical calculations of the time-dependent Schwinger-like pair production mechanism \cite{PhysRevLett.102.150404}. This is usually interpreted in terms of the Stokes phenomenon \cite{PhysRevLett.104.250402,PhysRevLett.108.030401} or Landau-Zener-St\"{u}ckelberg interferometry (LZSI) \cite{PhysRevA.86.032118} and makes the total rate sensitive to field parameters \cite{Abdukerim2013820}. Most known results have been obtained by comparing numerical methods to approximate analytical schemes such as semi-classical techniques \cite{PhysRevD.83.065028}, the worldline formalism \cite{PhysRevD.84.125023}, and the adiabatic-impulse model \cite{PhysRevA.86.032118}.  

This article focuses on the explanation of the two-dimensional momentum-space interference patterns in the electron momentum density induced by multiple avoided crossings of the adiabatic energies in graphene subjected to an oscillating electric field. 
 
Throughout, the Schwinger-like regime is considered where the dimensionless Keldysh parameter $\gamma$ obeys
\begin{eqnarray}
\label{eq:keldysh}
\gamma := \cfrac{m_{\perp}\omega v_{F}}{eE_{0}} \ll 1,
\end{eqnarray}
where $m_{\perp}:= \sqrt{\frac{\mathbf{p}_{\perp}^{2}}{v_{F}^{2}} + m_{\rm gap}^{2}}$ is the transverse mass, with $m_{\rm gap}$ the quasi-particle effective mass related to the gap $\Delta = m_{\rm gap}v_{F}^{2}$, $\mathbf{p}_{\perp}$ is the transverse momentum in a plane perpendicular to the external electric field, $e >0$ is the magnitude of the electron charge, $\omega$ is the frequency of the external field and $E_{0}$ is its electric field strength. The opposite case, where $\gamma \gg 1$, corresponds to the multiphoton regime and yields qualitatively different results \cite{PhysRevB.92.035401}.

This article is separated as follows. First, the pair production formalism in graphene is given in Sec. \ref{sec:can_quant}. Then, the adiabatic-impulse model and its relation to pair production is presented in Sec. \ref{sec:adia_model}. Numerical results for a simple oscillating field obtained from these two techniques are given and compared in Sec. \ref{sec:num_res}. In particular, the results for a few half-cycles can be found in Sec. \ref{sec:quant_inter}, where the concept of quantum interference is used to explain the qualitative differences in the electron momentum density for one and two half-cycles.
The long-time limit of the pair creation results is interpreted in terms of multiphoton quantum interference via Floquet theory in Sec. \ref{sec:long_time}. In Sec. \ref{sec:second_order}, we discuss the analogy between the adiabatic dynamics of quasiparticles in graphene and the Kibble-Zurek mechanism in second order phase transitions. Finally, the conclusion is in Sec. \ref{sec:conclu}.

\section{Pair production in a strong homogeneous field}
\label{sec:can_quant}

The formalism to compute the electron momentum density produced by a strong classical electromagnetic field is reviewed in Ref. \cite{Gelis20161} for QED. These QED techniques have been adapted to RQED$_{3,2}$ and applied to graphene physics: for more details, we refer the reader to the work presented in \cite{PhysRevA.86.032118,PhysRevB.92.035401}. Hereinafter, the main results of this analysis are given with an emphasis on the definition of quantities required to compute the electron density in graphene.  Other techniques to compute the pair density are also available \cite{PhysRevD.86.125022,PhysRevD.78.061701,PhysRevD.83.065007}.

It was demonstrated that the leading order contribution to the electron momentum density $d \langle \tilde{n}_{s,a} \rangle/d^{2}\mathbf{p}$ generated from electron-hole pair production for a graphene sample immersed in a homogeneous electric field can be written as \cite{PhysRevB.92.035401}
\begin{eqnarray}
\frac{d\langle \tilde{n}_{s,a}\rangle}{d^{2}\mathbf{p}}  =  
\frac{1}{2E_{\mathbf{p}}^{\rm out}2E_{\mathbf{p}}^{\rm in}}
 \left|  u^{\mathrm{out} \dagger}_{s,a}(\mathbf{p}) \psi_{s,a}(t_{f},\mathbf{p}) \right|^{2},
\label{eq:pair_prod_homo}
\end{eqnarray}
where $s=\pm 1$ denotes the physical spin of the electron and $a = \mathbf{K}_{\pm}$ indexes non-equivalent Dirac points. The wave function is given by 
\begin{eqnarray}
\label{eq:wf_init}
\psi_{s,a}(t_{f},\mathbf{p}) = U_{\mathbf{p}}(t_{f},t_{i}) v^{\mathrm{in}}_{s,a}(-\mathbf{p}),
\end{eqnarray}
where the evolution operator $U_{\mathbf{p}}$ evolves the initial wave function $v_{s,a}^{\mathrm{in}}$ from the initial asymptotic time $t_{i}$ to the final asymptotic time $t_{f}$ according to the following massless Dirac equation expressed in momentum space  \cite{gusynin2007ac}:
\begin{equation}\label{eq:dirac_eq_mom}
i\partial_{t}\psi_{s,\mathbf{K}_{\pm}}(t,\mathbf{p}) =  H_{\mathbf{K}_{\pm}}(t,\mathbf{p}) \psi_{s,\mathbf{K}_{\pm}}(t,\mathbf{p}),
\end{equation} 
with a Hamiltonian defined by
\begin{eqnarray}
\label{eq:hamiltonian}
H_{\mathbf{K}_{\pm}}(t,\mathbf{p}) :=  \pm v_{F}\boldsymbol{\sigma} \cdot \left[ \mathbf{p}  - q\mathbf{A}(t) \right] ,
\end{eqnarray} 
where $q$ is the electric charge ($q=-e$ for the electron), $\mathbf{A}$ is the time-dependent vector potential and $\boldsymbol{\sigma}$ are Pauli matrices. The electric field is given as usual by $\mathbf{E}(t) = - \partial_{t}\mathbf{A}(t)$. To derive Eq. \eqref{eq:pair_prod_homo}, it is assumed that the electric field vanishes at asymptotic times as $\left. \mathbf{E}(t)\right|_{t \in [-\infty,t_{i}] \cup [t_{f},\infty]} = 0$. Although the physical field is null in those asymptotic regions, it is possible that the vector potential has a constant value (the value depends on the gauge chosen). The constant value of the vector potential in these temporal regions will be denoted by $\left. \mathbf{A}(t)\right|_{t \in [-\infty,t_{i}]} = \mathbf{A}^{\mathrm{in}}$ and $\left. \mathbf{A}(t)\right|_{t \in [t_{f},\infty]} = \mathbf{A}^{\mathrm{out}}$.

Other parameters are also evaluated at asymptotic times in Eq. \eqref{eq:pair_prod_homo}. To define these parameters, it is convenient to introduce the kinematic momentum given by
\begin{eqnarray}
\mathbf{P}_{\pm}(t) := \pm \mathbf{p} -q \mathbf{A}(t).
\end{eqnarray}
Then, the adiabatic free spinors can be written as
\begin{eqnarray}
\label{eq:free_spin1}
u_{s,\mathbf{K}_{+}}(t,\mathbf{p}) &=& 
\cfrac{1}{\sqrt{E_{\mathbf{p}}(t)}}
\begin{bmatrix}
E_{\mathbf{p}}(t) \\
v_{F}\left[ P_{+,x}(t) + iP_{+,y}(t) \right]
\end{bmatrix}, \\ 
\label{eq:free_spin2}
v_{s,\mathbf{K}_{+}}(t,\mathbf{p}) &=& 
\cfrac{1}{\sqrt{E_{-\mathbf{p}}(t)}}
\begin{bmatrix}
v_{F}\left[-P_{-,x}(t) + iP_{-,y}(t) \right]\\
E_{\mathbf{-p}}(t) 
\end{bmatrix} ,\\
\label{eq:free_spin3}
u_{s,\mathbf{K}_{-}}(t,\mathbf{p}) &=& \cfrac{1}{\sqrt{E_{\mathbf{p}}(t)}}
\begin{bmatrix}
E_{\mathbf{p}}(t) \\
v_{F}\left[-P_{+,x}(t) - iP_{+,y}(t)\right]
\end{bmatrix} ,\\
\label{eq:free_spin4}
v_{s,\mathbf{K}_{-}}(t,\mathbf{p}) &=& \cfrac{1}{\sqrt{E_{\mathbf{-p}}(t)}}
\begin{bmatrix}
v_{F}\left[P_{-,x}(t) - iP_{-,y}(t) \right]\\
E_{\mathbf{-p}}(t) 
\end{bmatrix} ,
\end{eqnarray}
where the energy is defined as
\begin{eqnarray}
E_{\pm \mathbf{p}}(t) &:=& v_{F}| \mathbf{P}_{\pm}(t) |.
\end{eqnarray}
The spinors obey the usual property $u^{\dagger}_{s,a}(t,\mathbf{p})v_{s,a}(t,-\mathbf{p}) = 0$. In Eqs. \eqref{eq:pair_prod_homo} and \eqref{eq:wf_init}, free spinors have a subscript $\mathrm{in/out}$, denoting that these spinors are evaluated at times $t_{i}$ and $t_{f}$, respectively ($u_{s,a}^{\mathrm{out}}(\mathbf{p}) := u_{s,a}(t_{f},\mathbf{p})$ and $v_{s,a}^{\mathrm{in}}(\mathbf{p}): = v_{s,a}(t_{i},\mathbf{p})$).

To summarize, the electron momentum density is computed by preparing a free negative energy state with momentum $\mathbf{p}$ at time $t_{i}$, by evolving this state up to the final time $t_{f}$ with the Dirac equation coupled to the field, and by projecting this final state on a free positive energy state $u_{s,a}^{\mathrm{out}}$. This procedure is performed for all momenta. The time evolution can be computed by resorting to analytical solutions of the Dirac equation or by employing a numerical scheme. The latter option is taken here where a split-operator decomposition of the evolution operator developed in previous studies \cite{PhysRevA.86.032118,PhysRevB.92.035401} is utilized.

\section{Electron-hole production in the adiabatic-impulse model}
\label{sec:adia_model}

In a homogeneous electric field, pair production is computed by solving Eq. \eqref{eq:dirac_eq_mom}, which is analogous to a quantum two-level system \cite{PhysRevA.86.032118}. As a consequence, many of the analytical techniques developed to study this class of quantum systems can be employed to evaluate the pair or electron momentum density. These approaches are important to understand the physics of pair creation in some given regime. In particular, we are interested in the adiabatic limit, characterized by the following condition \cite{PhysRevA.55.R2495}:
\begin{eqnarray}
\Omega  \ll  \min \left\{ ev_{F} \max_{t\in \mathbb{R}}|\mathbf{A}(t)|, v_{F} |\mathbf{p}_{\perp}|\right\},
\end{eqnarray} 
where $\Omega$ is the characteristic inverse time scale for the variation of the electromagnetic potential, assuming that the latter can be written as $\mathbf{A}(\Omega t)$.

In this regime, the quantum two-level system has been studied extensively within the adiabatic perturbation theory formalism \cite{zener1932non,stueckelberg1932two,Shevchenko20101,PhysRevA.55.4418,nakamura:4031,PhysRevA.73.063405,PhysRevA.55.R2495, Berry61,doi:10.1080/00268977000101041,PhysRevA.75.063414,Suqing2005315}. In this adiabatic limit, it has been demonstrated that the quantum dynamics proceeds by a sequence of adiabatic evolution followed by non-adiabatic transitions. This can be approximated through the adiabatic-impulse model, which is now used to compute the electron density and to obtain an intuitive understanding of the interference phenomenon occurring in pair production.

The wave function in the adiabatic basis can be expressed as
\begin{eqnarray}
\psi_{s,a}(t,\mathbf{p}) &=& B^{(u)}_{s,a}(t)u_{s,a}(t,\mathbf{p}) + B^{(v)}_{s,a}(t)v_{s,a}(t,-\mathbf{p}),
\end{eqnarray}
where $B^{(u,v)}_{s,a}$ are time-dependent coefficients of the adiabatic basis expansion.
Using the properties of free spinors, the electron momentum density of Eq. \eqref{eq:pair_prod_homo}, in the adiabatic approximation, is written as
\begin{eqnarray}
\frac{d\langle \tilde{n}_{s,a} \rangle}{d^{2}\mathbf{p}} =   
\frac{E_{\mathbf{p}}^{\rm out}}{E_{\mathbf{p}}^{\rm in}}
 \left| B^{(u)}_{s,a}(t_{f}) \right|^{2}.
\label{eq:pair_prod_adia}
\end{eqnarray}
Here, the initial condition in the adiabatic basis is given by $ \mathbf{B}_{s,a}(t_{i}) = [0,1]^{\mathrm{T}}$. This is consistent with the required initial condition for the pair density calculation given in Eq. \eqref{eq:wf_init}.

The result in Eq. \eqref{eq:pair_prod_adia} is independent of the representation of Dirac matrices in which the free spinors $u,v$ and the wave function are expressed. Therefore, in the following discussion, it is assumed that $u,v$ are the solutions of the Dirac equation in Eq. \eqref{eq:dirac_eq_mom} with the substitutions
\begin{eqnarray}
\sigma_{x} \rightarrow \sigma_{z} \; \mbox{ and} \; \sigma_{y} \rightarrow \sigma_{x}.
\end{eqnarray}
This change of representation can be performed via the unitary transformation
\begin{eqnarray}
\label{eq:unitary}
U_{r}:=e^{-i\sigma_{y}\frac{\pi}{4}} e^{-i\sigma_{x}\frac{\pi}{4}}.
\end{eqnarray}  
These transformations allow for the direct application of the adiabatic-impulse results given in Ref. \cite{Shevchenko20101}. The latter is now discussed for a more general field time-dependence. In particular, we consider an homogeneous electric field linearly polarized in the $x$-direction. In this case, the vector potential has only one non-zero component and can be written as
\begin{eqnarray}
\mathbf{A}(t) = 
\begin{bmatrix}
A_{0}+A(t) \\ 0
\end{bmatrix}
,
\end{eqnarray} 
where $A_{0}$ is a constant shift of the vector potential while $A(t)$ is the function that determines its time dependence. The general case, where all the components are non-zero, can also be handled in principle and gives rise to the well-known Berry phase \cite{Berry61}. This however is outside the scope of this article.

\begin{figure}
\includegraphics[width=0.5\textwidth]{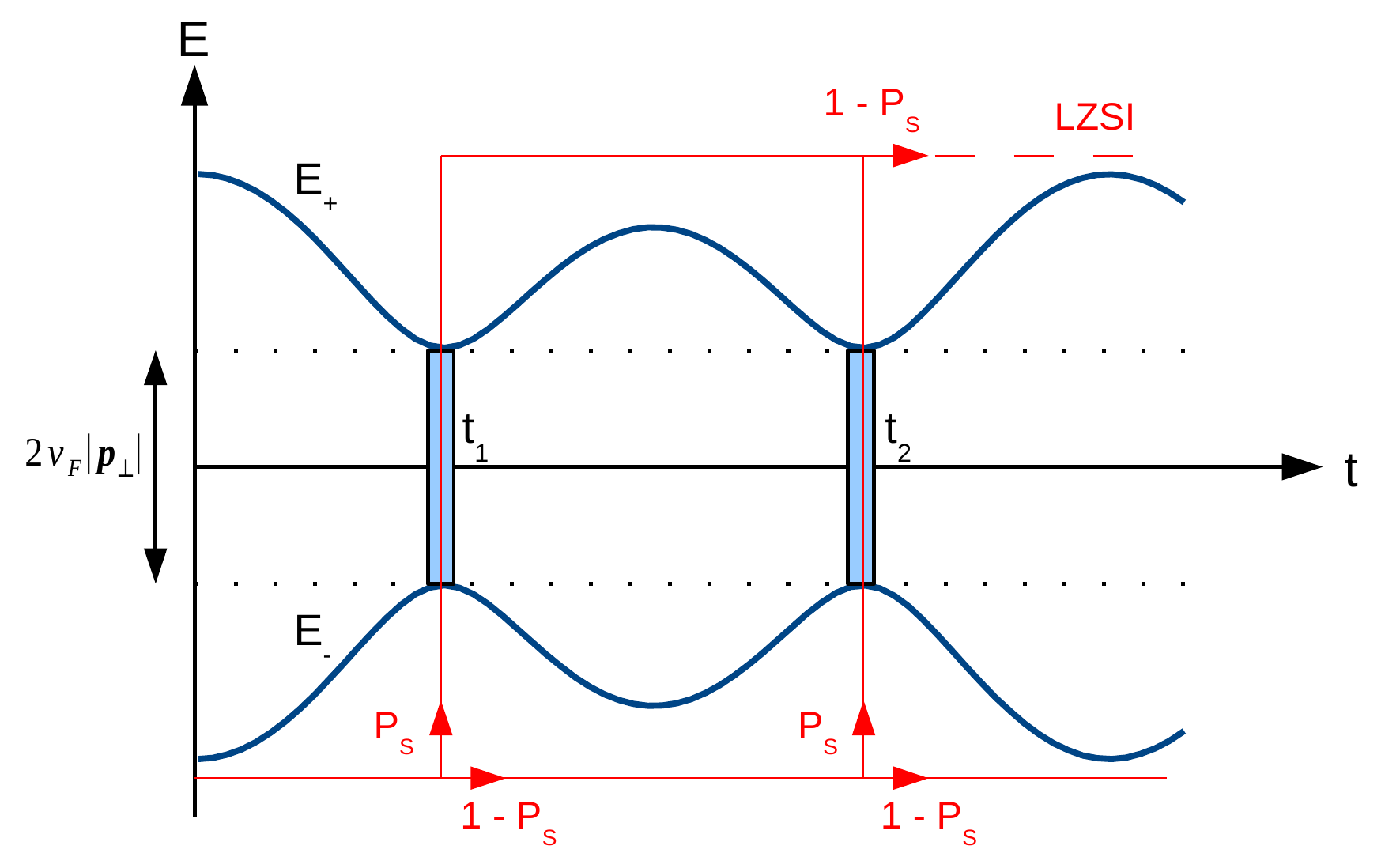}
\caption{Adiabatic energies in the driven two-level model. The adiabatic energies are denoted by $E_{\pm}$ while $P_{S}$ represents the transition probability. Nonadiabatic transitions occur at times $t_{1,2,\cdots}$ where the gap is minimal and given by $ \Delta(t_{1,2}) = 2v_{F}|\mathbf{p}_{\perp}|$. The red lines represent different transition paths from negative to positive energy states. At time $t_{2}$ and all other times afterwards where a nonadiabatic transition takes place, the negative energy portion of the wave function that transits upward with probability $P_{S}$ interferes with the positive energy part. This is the LZSI.}
\label{fig:adiabatic}
\end{figure}

For this class of potential, the coefficient $B^{(u)}$ can be determined in the adiabatic-impulse model where the quantum dynamics proceeds in steps where adiabatic evolutions are followed by non-adiabatic transitions. Defining a vector in the adiabatic basis space as $\mathbf{B}_{s,a}(t):=(B^{(u)}_{s,a}(t),B^{(v)}_{s,a}(t))^{\mathrm{T}}$, the time evolution of the adiabatic coefficients can then be given as \cite{Shevchenko20101}
\begin{eqnarray}
\label{eq:adiab_evol}
\mathbf{B}_{s,a}(t_{f}) &=& U_{\mathrm{adia}}(t_{f},t_{n})N_{n} U_{\mathrm{adia}}(t_{n},t_{n-1})N_{n-1} \nonumber \\
&& \cdots N_{2}U_{\mathrm{adia}}(t_{2},t_{1})N_{1} U_{\mathrm{adia}}(t_{1},t_{i}) \mathbf{B}_{s,a}(t_{i}), \nonumber \\
\end{eqnarray}
where $t_{1},\cdots,t_{n}$ are times when there is an avoided crossing and when non-adiabatic transitions take place, as depicted in Fig. \ref{fig:adiabatic}. These occur at complex times $(t^{*}_{j})_{j=1,\cdots,n}$ when $E_{\mathbf{p}}(t^{*}_{j}) = -E_{\mathbf{p}}(t^{*}_{j}) = 0$, i.e. when the positive and negative adiabatic energies cross in the complex time plane \cite{Shevchenko20101,PhysRevA.44.4280}. The crossing times are then given by $t_{j} = \mathrm{Re}(t^{*}_{j})|_{j=1,\cdots,n}$. For $\gamma \ll 1$ and according to the adiabatic-impulse model, these times can be evaluated approximately by determining when the adiabatic mass gap $\Delta(t) := 2E_{\mathbf{p}}(t)$ is minimal, as shown in Appendix \ref{app:min_gap}. This also requires the squared canonical momentum to be minimal. Therefore, the crossing times are solutions of the following minimization problem (for $j=1,\cdots,n$):
\begin{eqnarray}
t_{j} = \min_{t\in T_{j}}P_{+,x}^{2}(t) = \min_{t\in T_{j}}\left[ p_{x} + eA_{0} + e A(t) \right]^{2},
\end{eqnarray} 
where $T_{j}$ represents the $j$th time interval where the function is convex. This minimization problem can be solved by computing the first and second time derivatives of $P_{+,x}^{2}(t)$. Therefore, the transition times are solutions of the following system of equations:
\begin{eqnarray}
\label{eq:cond1}
 \left[  p_{x} + eA_{0} + e A(t) \right]E_{x}(t) &=& 0, \\
 \label{eq:cond2}
 - \left[  p_{x} + eA_{0} + e A(t) \right] \partial_{t}E_{x}(t)  + E_{x}^{2}(t) &>& 0,
\end{eqnarray}
obtained from the first and second time derivatives, respectively. Equations \eqref{eq:cond1} and \eqref{eq:cond2} yield two independent cases:
\begin{eqnarray}
\label{eq:case1}
&&
\begin{cases}
P_{+,x}(t) := p_{x} + eA_{0} + e A(t) = 0,\\
E_{x}^{2}(t) > 0,
\end{cases} \\
\label{eq:case2}
&&
\begin{cases}
E_{x}(t) = 0,\\
-P_{+,x}(t) \partial_{t}E_{x}(t) > 0.
\end{cases}
\end{eqnarray}
In the first case (Eq. \eqref{eq:case1}), the inequality is always fulfilled because $E^{2}_{x}$ is positive-definite. Accordingly, solutions of $P_{+,x}(t) = 0$ provide times where the mass gap is minimal and given by $\Delta(t_{j}) = 2v_{F}|\mathbf{p}_{\perp}|$. However, there may exist values of $p_{x}$ for which $P_{+,x}(t) = 0$ has no solution. These occurrences are covered by the second case in Eq. \eqref{eq:case2}, which corresponds physically to a vanishing electric field. In this case, minima are found when $P_{+,x}(t)$ and $\partial_{t}E_{x}(t)$ have opposite signs. When the electric field is zero, the nonadiabatic transition probability is vanishing and therefore, the time evolution is adiabatic. As a consequence, the adiabatic-impulse approach predicts that there is no contribution to the electron momentum density from these times because transitions are forbidden. Henceforth, we will only consider the first case given in Eq. \eqref{eq:case1}, assuming that no pairs are produced when the second case (Eq. \eqref{eq:case2}) is fulfilled. As seen below in numerical results, the model is not accurate in this latter case. The reason for this discrepancy can be traced back to the fact that in this regime, the times when there is a minimal gap do not correspond to times where the adiabatic energies are crossings, as discussed in Appendix \ref{app:min_gap}. This is a limitation of the adiabatic-impulse model.

In Eq. \eqref{eq:adiab_evol}, the operator $U_{\mathrm{adia}}$ is the adiabatic evolution operator given by 
\begin{eqnarray}
U_{\mathrm{adia}}(t_{n-1},t_{n}) &:=& \exp \left[ -i \sigma_{z} \int_{t_{n-1}}^{t_{n}} E_{+\mathbf{p}}(t)dt \right], \\
&=& \exp \left[ -i \sigma_{z} \xi_{n} \right],
\end{eqnarray}
where $\xi_{n}$ is the accumulated phase in the adiabatic evolution of the system. 
Conversely, the matrices $N_{j}|_{j=1,\cdots,n}$ are obtained by solving the Dirac equation in a time region close to $t_{j}$ by shifting to $t'$. Close to these times, the potential is linearized and expressed as
\begin{eqnarray}
A(t_{j}+t') & \approx & A(t_{j}) + t' \partial_{t}A(t)|_{t=t_{j}}, 
\end{eqnarray}
where $t'$ is some small time. When the first condition in Eq. \eqref{eq:case1} is fulfilled, the last equation can be written as
\begin{eqnarray}
\label{eq:lin_potential}
A(t_{j}+t') &\approx &  -\frac{p_{x}}{e} - A_{0} - t'E_{x}(t_{j}).
\end{eqnarray}

Then, using Eq. \eqref{eq:lin_potential}, the Dirac equation (Eq. \eqref{eq:dirac_eq_mom}) becomes formally similar to the Landau-Zener problem, which can be solved exactly using parabolic cylinder functions \cite{zener1932non}. Matching this solution to the adiabatic solution using the asymptotic expansion of parabolic cylinder functions, it is possible to determine a transition matrix. It is given by \cite{nakamura:4031,Shevchenko20101}:
\begin{eqnarray}
\label{eq:N}
 N_{j} :=
\begin{bmatrix}
 \sqrt{1-P^{(j)}_{S}(\mathbf{p})}e^{-i \tilde{\phi}_{j} } & -\sqrt{P^{(j)}_{S}(\mathbf{p})} \\ 
 \sqrt{P^{(j)}_{S}(\mathbf{p})} & \sqrt{1-P^{(j)}_{S}(\mathbf{p})}e^{i \tilde{\phi}_{j} }
\end{bmatrix},
\end{eqnarray}
where the Stokes phase, characterizing the phase accumulated during nonadiabatic transitions, is defined as
\begin{eqnarray}
\label{eq:stokes_ph}
 \tilde{\phi}_{j} := - \frac{\pi}{4} + \delta_{j} [\ln (\delta_{j}) -1] +  \arg \Gamma(1-i\delta_{j}), 
\end{eqnarray}
with
\begin{eqnarray}
\label{eq:delta}
\delta_{j}:= \frac{v_{F}p_{y}^{2}}{2e |E_{x}(t_{j})|}.
\end{eqnarray}
The transition probability is then
\begin{eqnarray}
\label{eq:landau_zener}
P^{(j)}_{S}(\mathbf{p}) = e^{-2\pi \delta_{j}}. 
\end{eqnarray} 

We are now in a position to consider a few cases of interest. When there is one avoided crossing, the electron momentum density is given by
\begin{eqnarray}
\frac{d\langle \tilde{n}_{s,a} \rangle}{d^{2}\mathbf{p}} =   
 P^{(j)}_{S}(\mathbf{p}).
\label{eq:pair_prod_adia_one}
\end{eqnarray}
On the other hand, when there are two avoided crossings, it can be shown that 
\begin{eqnarray}
\frac{d\langle \tilde{n}_{s,a} \rangle}{d^{2}\mathbf{p}} &=&   
\biggl[ P^{(1)}_{S}(\mathbf{p})  + P^{(2)}_{S}(\mathbf{p}) - 2P^{(1)}_{S}(\mathbf{p})P^{(2)}_{S}(\mathbf{p})\nonumber \\
 && + \sqrt{P^{(1)}_{S}(\mathbf{p})P^{(2)}_{S}(\mathbf{p}) [1-P^{(1)}_{S}(\mathbf{p})][1-P^{(2)}_{S}(\mathbf{p})]} \nonumber \\
 && \quad \quad \times \cos(2\xi_{2} + \tilde{\phi}_{1}+ \tilde{\phi}_{2})\biggr].
\label{eq:pair_prod_adia_two}
\end{eqnarray}
Other relations exist for any number of crossings but are not shown here for simplicity.

\section{Numerical results and discussion}
\label{sec:num_res}

In this section, numerical results are obtained using the computational techniques described in Sec. \ref{sec:can_quant} and \ref{sec:adia_model}. In the long time limit, Floquet theory is also introduced to explain some general features of the electron momentum density. 

A simple homogeneous oscillating field is considered. The latter can be generated experimentally by using counterpropagating laser fields where the magnetic field is cancelled. The electric field is characterized by ($n\in \mathbb{N}^{+}$ is the number of half-cycles)
\begin{eqnarray}
\label{eq:oscill_field}
E_{x}(t) &=& 
\begin{cases}
0 & \mbox{for} \;\; t<0 \\
E_{0}\sin(\omega t)& \mbox{for} \;\; t \in [0,n\pi/\omega] \\
0 & \mbox{for} \;\; t>n\pi/\omega 
\end{cases} \;\; , \\
A(t) &=& \label{eq:vector_potential}
\begin{cases}
\frac{E_{0}}{\omega} & \mbox{for} \;\; t<0 \\
\frac{E_{0}}{\omega}\cos(\omega t) & \mbox{for} \;\; t \in [0,n\pi/\omega] \\
\begin{cases}
\frac{E_{0}}{\omega} & n \;\; \mbox{even} \\
-\frac{E_{0}}{\omega} & n \;\; \mbox{odd}
\end{cases}
 & \mbox{for} \;\; t>n\pi/\omega 
\end{cases}  \;\;,\\
A_{0} &=& 
\begin{cases}
-\frac{E_{0}}{\omega} & n \;\; \mbox{even} \\
\frac{E_{0}}{\omega} & n \;\; \mbox{odd}
\end{cases}.
\end{eqnarray}
For this electric field, the electron momentum density at zero transverse momentum $p_{y} = 0$ can be evaluated analytically \cite{PhysRevB.92.035401}. For $n$ even, the electron momentum density is zero while for $n$ odd, it is given by
\begin{eqnarray}
\label{eq:rate_oscil_pyis0}
 \left. \cfrac{d\langle \tilde{n}_{s,\mathbf{K}_{\pm}} \rangle}{d^{2}\mathbf{p}} \right|_{p_{y} = 0} =   
\begin{cases}
0 & \mbox{if} \;\; p_{x} < -e\frac{2E_{0}}{\omega}\\
0 & \mbox{if} \;\; p_{x} > 0 \\
1 & \mbox{if} \;\; p_{x} < 0 \;\; \mbox{and} \;\; p_{x} > -e\frac{2E_{0}}{\omega}
\end{cases}.
\end{eqnarray}
%

\subsection{Quantum interferences}
\label{sec:quant_inter}

\begin{figure}
\subfloat[]{\includegraphics[width=0.5\textwidth]{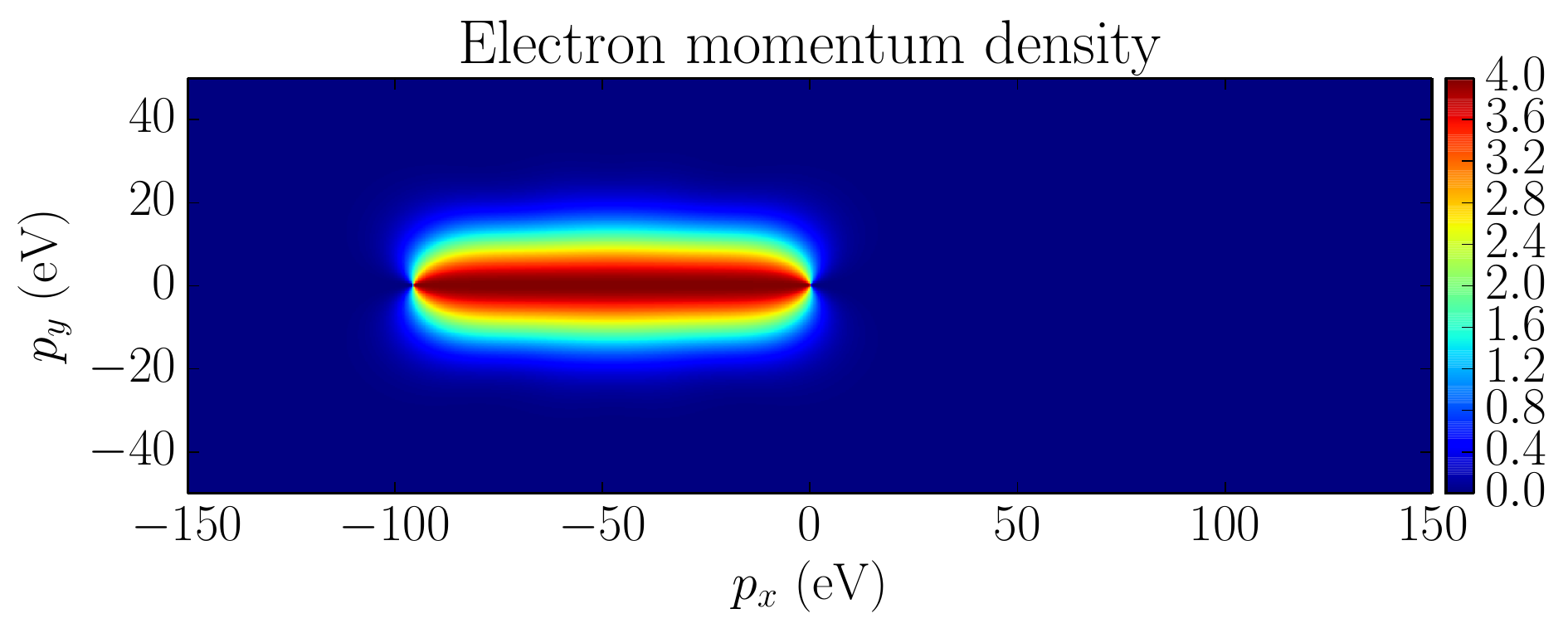}}\\
\subfloat[]{\includegraphics[width=0.5\textwidth]{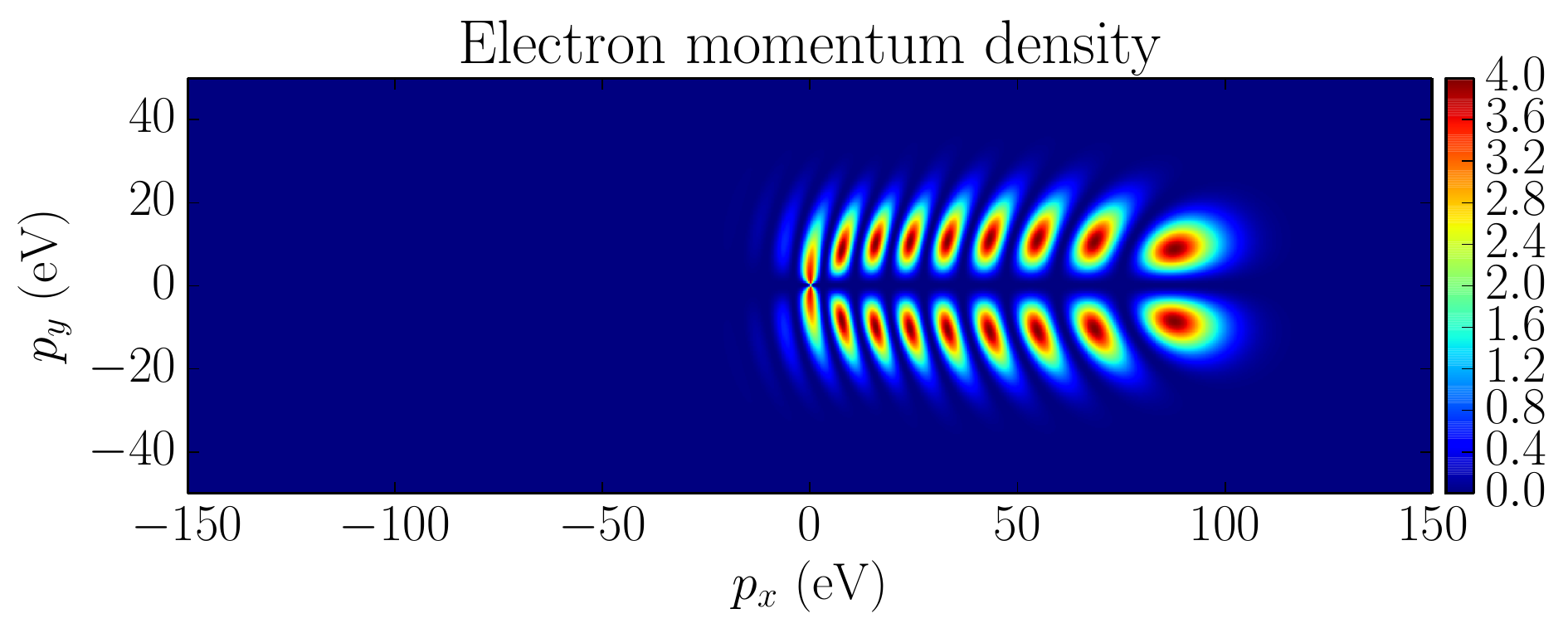}} 
\caption{Numerical results for the electron momentum density for an oscillating external field linearly polarized in the $x$-coordinate, with a field strength of $E_{0} = 1.0 \times 10^{7}$ V/m and a frequency of $\nu = 10.0$ THz. (a) Electron momentum density after a semi-cycle. (b) Electron momentum density after a full cycle. An interference pattern can be seen in (b) where a peak and valley structure appears in the electron momentum density.}
\label{fig:density_inter}
\end{figure}

The numerical results for the electron momentum density produced by an electric field of strength $E_{0} = 1.0 \times 10^{7}$ V/m and frequency $\nu = 10.0$ THz are displayed in Fig. \ref{fig:density_inter} for a half-cycle ($n=1$) and for a full cycle ($n=2$). These results are obtained using the numerical technique presented in Sec. \ref{sec:can_quant}. When the field is applied for a half-cycle, no interference pattern can be observed, as seen in Fig. \ref{fig:density_inter} (a). In this case, the electron momentum density is maximal at $p_{y}=0$ and is non-zero on the momentum interval $p_{x} \in [-95.4 \ \mbox{eV},0]$, consistent with Eq. \eqref{eq:rate_oscil_pyis0}.

On the other hand, when the field is applied for a full cycle, the electron momentum density reveals large variations of the density over the momentum range considered, shifting from zero density to values close to $\approx 4.0$, the largest value allowed by the exclusion principle. This is typical of an interference pattern: it induces a ``peak and valley structure'' where the electron momentum density oscillates rapidly over the momentum domain. As explained in more details in the following, this can be interpreted as time domain quantum interference and is an example of Landau-Zener-St\"{u}ckelberg interferometry.

\begin{figure}
\subfloat[]{\includegraphics[width=0.5\textwidth]{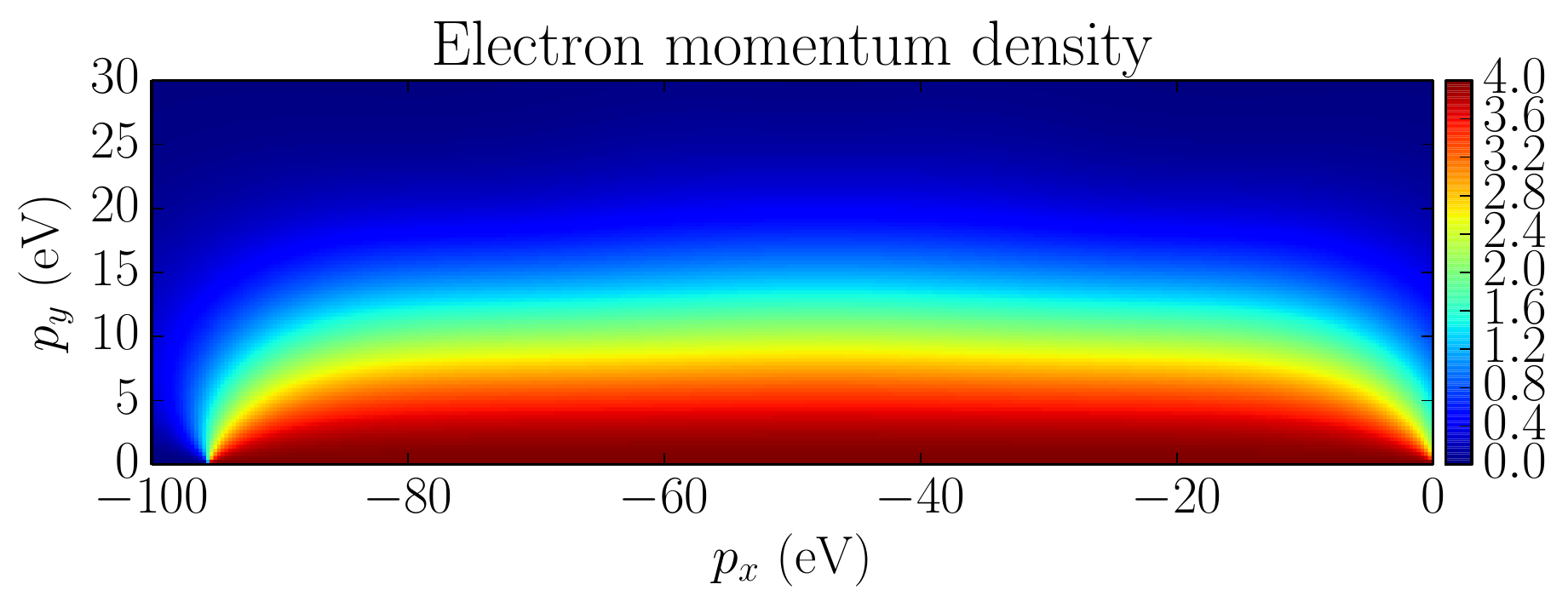}} \\
\subfloat[]{\includegraphics[width=0.5\textwidth]{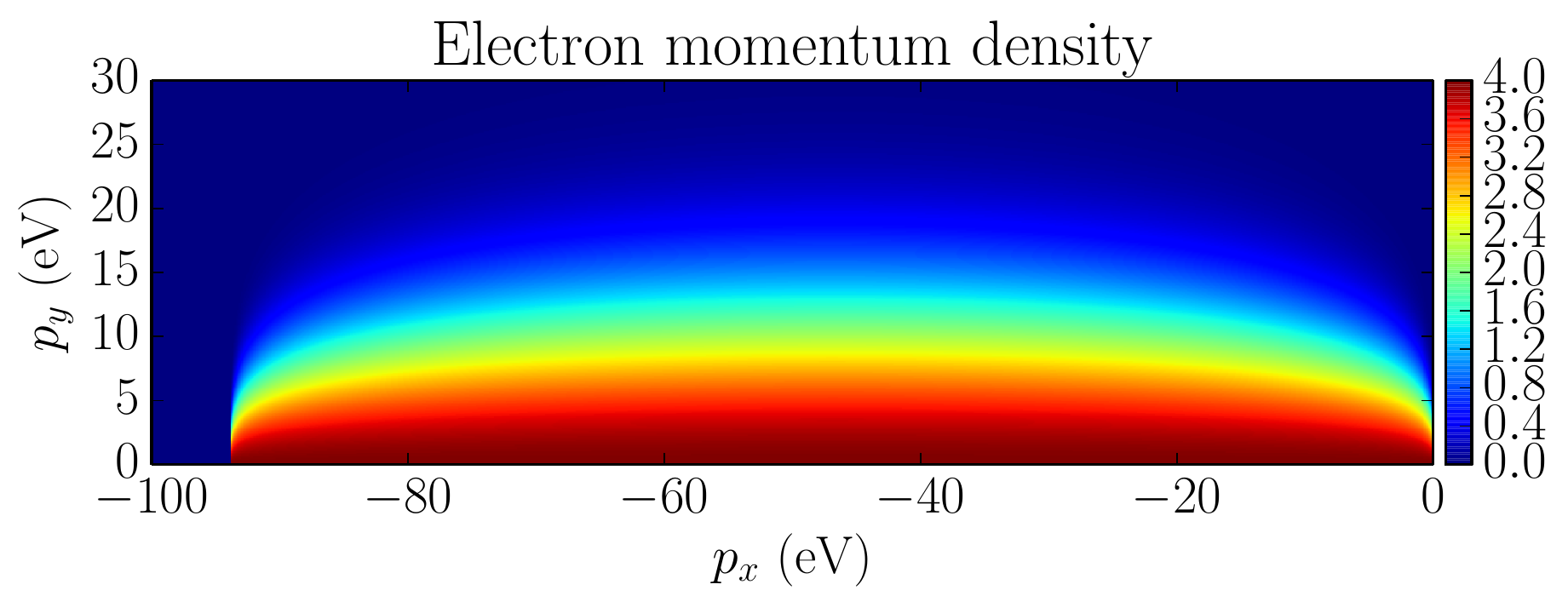}}
\caption{Comparison between the full numerical approach (a) and the adiabatic-impulse model (b) for the calculation of the electron momentum density. The electric field considered is linearly polarized in the $x$-coordinate and has a field strength of $E_{0} = 1.0 \times 10^{7}$ V/m, a frequency of $\nu = 10.0$ THz and is applied for a half-cycle.}
\label{fig:adia_vs_full_half}
\end{figure}

\begin{figure}
\subfloat[]{\includegraphics[width=0.5\textwidth]{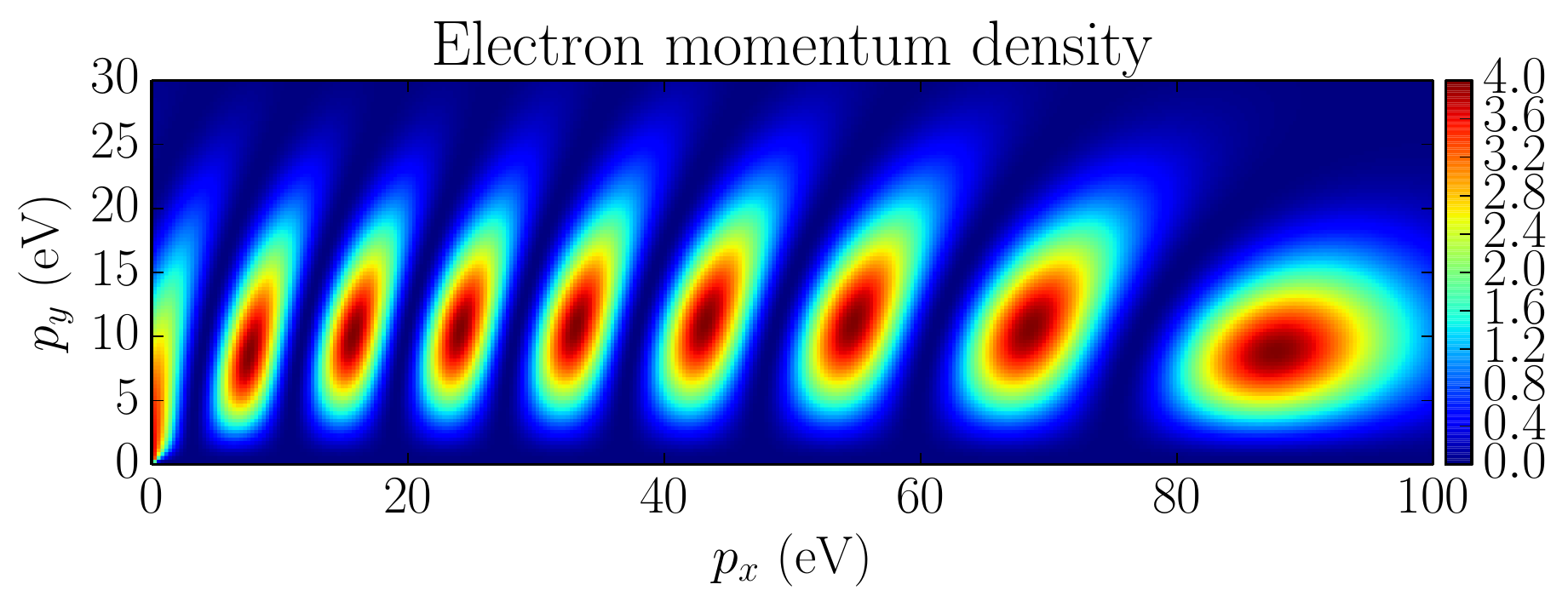}} \\
\subfloat[]{\includegraphics[width=0.5\textwidth]{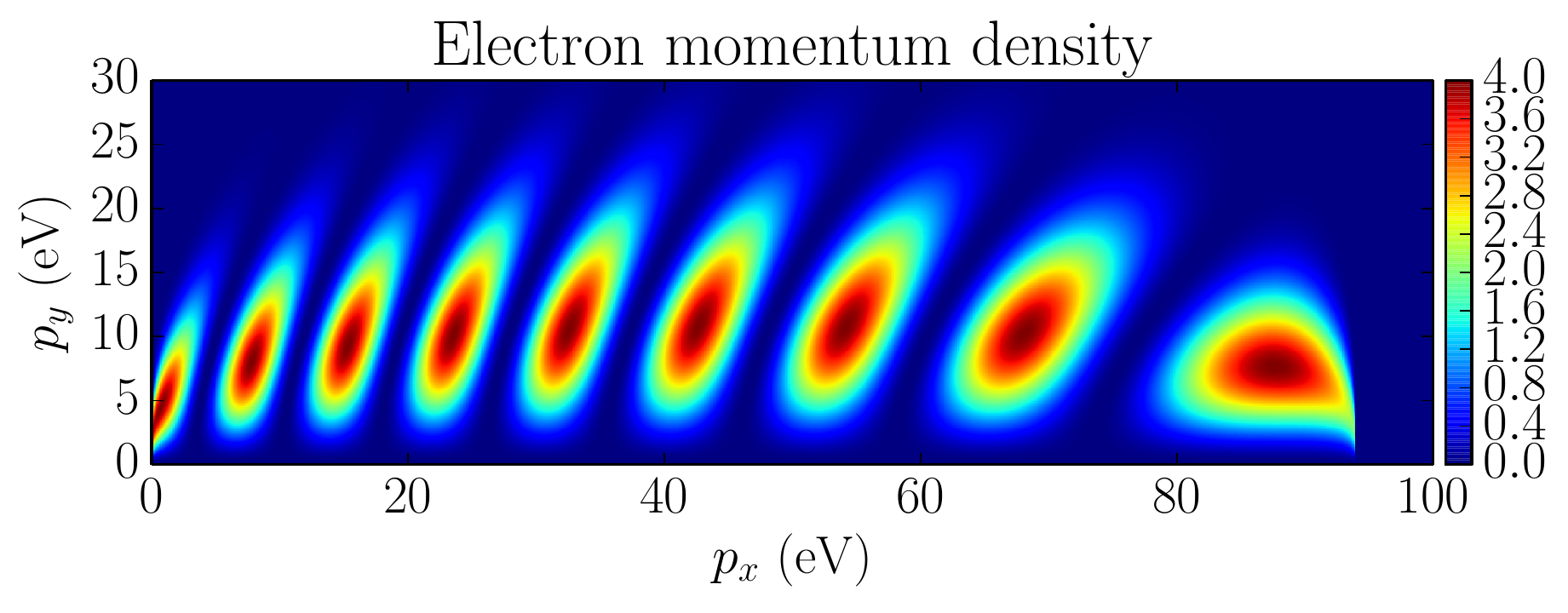}}
\caption{Comparison between the full numerical approach (a) and the adiabatic-impulse model (b) for the calculation of the electron momentum density. The electric field considered is linearly polarized in the $x$-coordinate and has a field strength of $E_{0} = 1.0 \times 10^{7}$ V/m, a frequency of $\nu = 10.0$ THz and is applied for a full cycle.}
\label{fig:adia_vs_full_one}
\end{figure}

The numerical results in Fig. \ref{fig:density_inter} are consistent with the ones obtained from the adiabatic-impulse model described in Sec. \ref{sec:adia_model}, as displayed in Figs. \ref{fig:adia_vs_full_half} and \ref{fig:adia_vs_full_one}. Both approaches yield an electron momentum density qualitatively similar, having maxima and minima at the same momenta. Using the intuitive physical interpretation of the adiabatic-impulse model, it can be concluded that pair production in graphene, in the adiabatic regime, occurs by a sequence of adiabatic evolutions followed by nonadiabatic transitions arising when the energy gap is minimal. The interference pattern appears after one cycle, as seen in Fig. \ref{fig:adia_vs_full_one}, because the lower and upper energy states accumulate different phases. Then, these energy states are coherently recombined at each nonadiabatic transition, resulting in quantum interference patterns for $n>1$. When they interfere constructively (destructively), the result is a maxima (minima) in the electron momentum density. This makes for a realization of Landau-Zener-St\"{u}ckelberg interferometry (defined in Fig. \ref{fig:adiabatic}) using quasiparticles in graphene. 

By comparing numerical results for the half-cycle (Fig. \ref{fig:adia_vs_full_half}) with the adiabatic-impulse model, one can also explain the directionality of the electron momentum density. The transition probability, given in Eq. \eqref{eq:landau_zener}, is exponentially suppressed at higher transverse momenta, confirming that the transverse momentum acts like a mass gap since $P_{S}^{(j)}(\mathbf{p})$ has the same form as the Schwinger probability in Eq. \eqref{eq:proba}. 

\subsection{Long time limit: Floquet theory}
\label{sec:long_time}

\begin{figure*}
\subfloat[]{\includegraphics[width=0.7\textwidth]{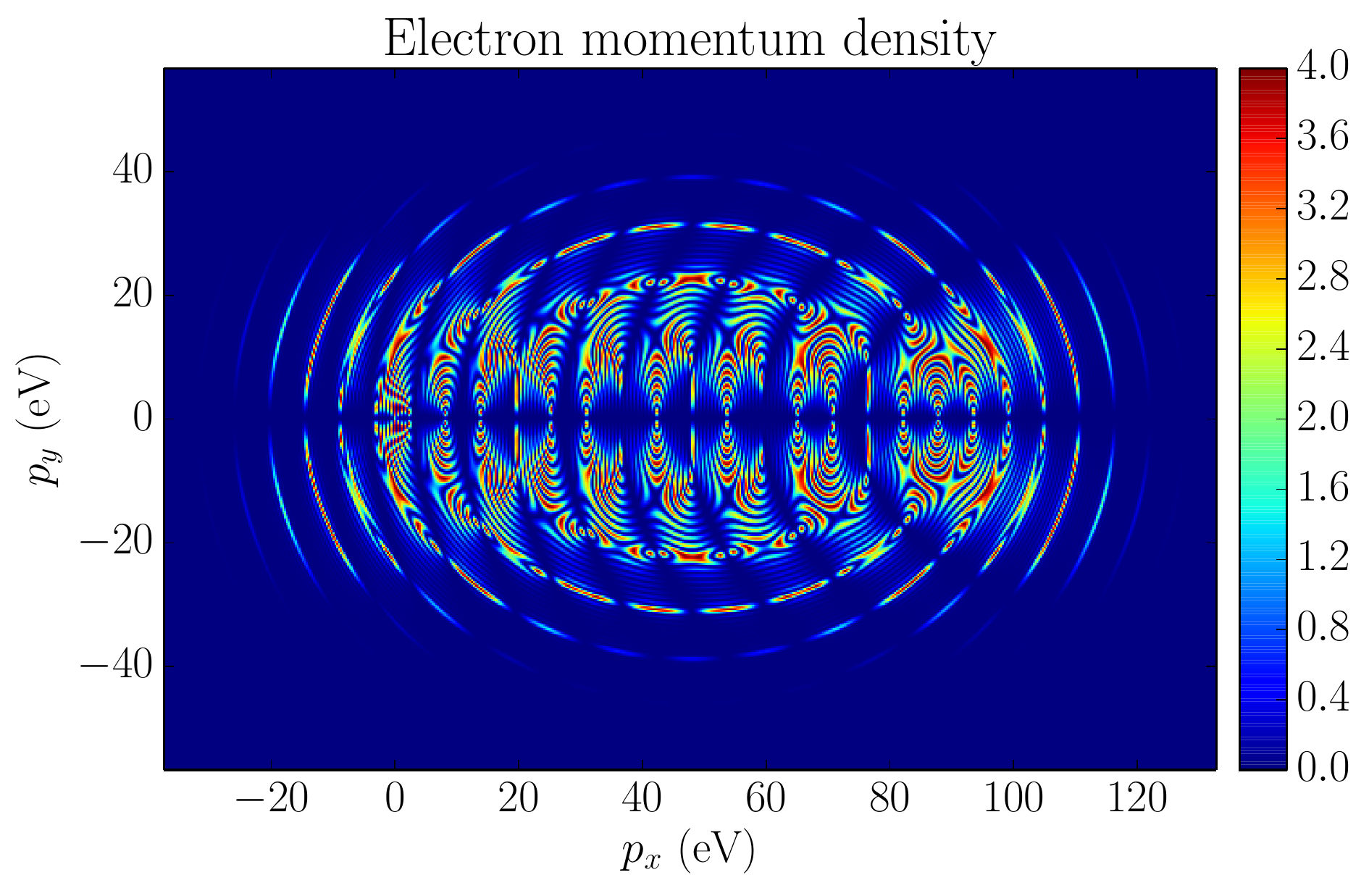}}\\
\subfloat[]{\includegraphics[width=0.7\textwidth]{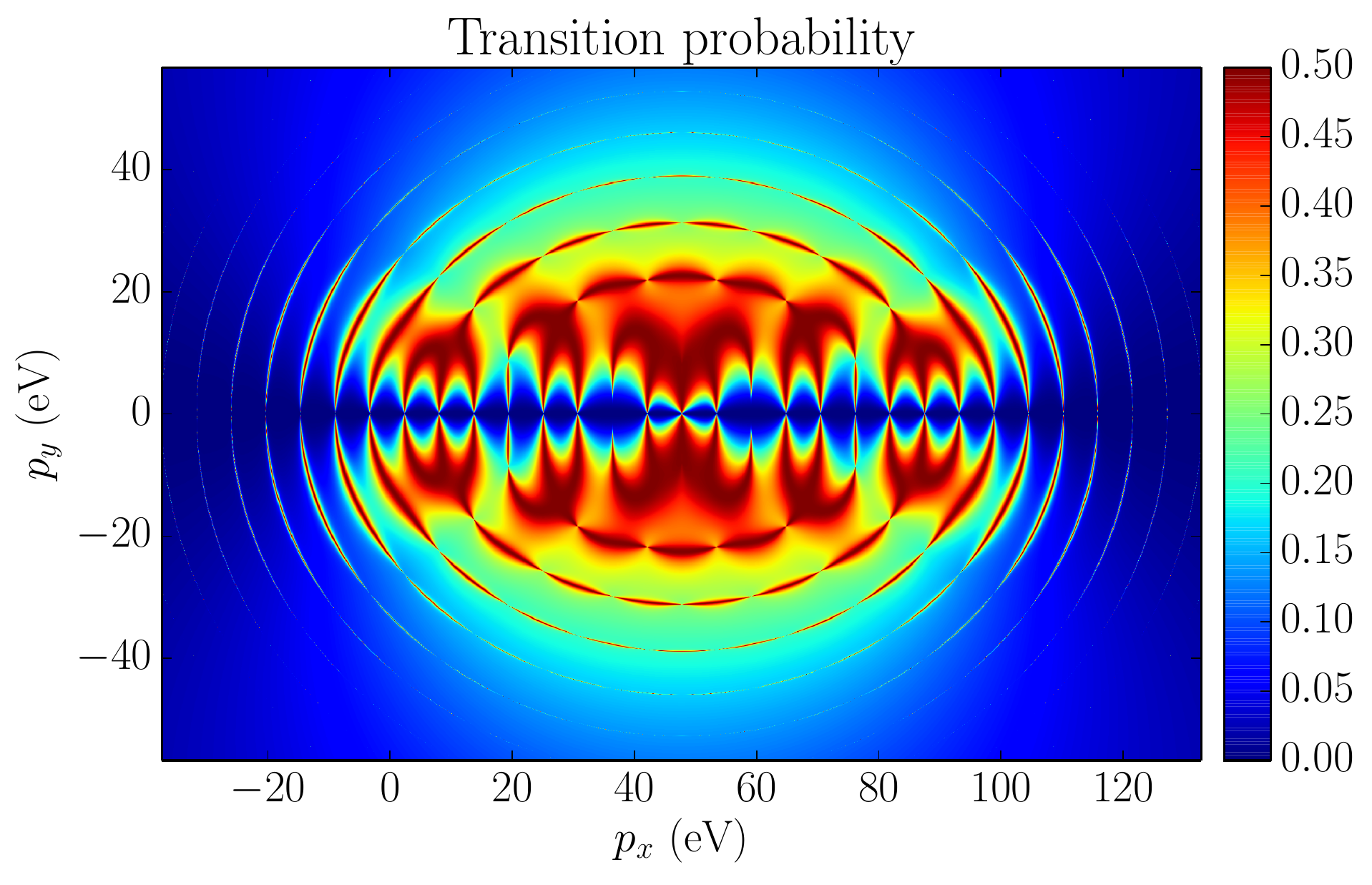}}
\caption{ (a) Numerical results for the electron momentum density for an oscillating external field linearly polarized in the $x$-coordinate, with a field strength of $E_{0} = 1.0 \times 10^{7}$ V/m and a frequency of $\nu = 10.0$ THz. The electron momentum density is calculated in the long time limit, after ten cycles. An interference pattern can be seen where an intricate peak and valley structure appears in the electron momentum density. (b) Time-averaged transition probability between field-free eigenstates $\ket{-}$ and $\ket{+}$ computed via Floquet theory (Eq. \ref{eq:probability}) for the same driving field.
	}
\label{fig:density_long}
\end{figure*}

The numerical results in the long time limit, after ten cycles ($n=20$), are displayed in Fig. \ref{fig:density_long}(a). The electron momentum density forms an intricate pattern where fast oscillations are superimposed over slowly varying and ring-like structures. The fast oscillations originate from the accumulated adiabatic phase and quantum interference, as in the one cycle case discussed in the last section. For $n \gg 1$, the system goes through many nonadiabatic transitions and therefore, there are several possible paths generating a transition from negative to positive energy states. For each path, a different phase is accumulated resulting in constructive and destructive interferences. This produces fast oscillations in the electron momentum density.

The other slowly-varying structures presented in Fig. \ref{fig:density_long}(a) can be explained in term of multiphoton quantum interference via Floquet theory.
The Floquet treatment side-steps the need for exhaustive time-dependent calculations, instead requiring the diagonalization of the Floquet Hamiltonian, an infinite-dimensional time-independent matrix.
Starting from the Dirac equation, Eq. \eqref{eq:dirac_eq_mom}, using the unitary transformation in Eq. \eqref{eq:unitary} and the vector potential in Eq. \eqref{eq:vector_potential} for an even number of cycles, one can write the graphene Hamiltonian as
\begin{equation}\label{eq:two_level}
H_{\mathbf{K}_{\pm}}(t,\mathbf{p}) =
-\frac{1}{2} (\varepsilon_x + \mathcal{A} \cos \omega t) \sigma_z - \frac{1}{2} \varepsilon_y \sigma_x,
\end{equation}
where
\begin{align}
\varepsilon_x & := \mp 2 v_F p_x - \mathcal{A}, \\
\varepsilon_y & := \mp 2 v_F p_y ,\\
\mathcal{A} &:=  \mp 2 v_F  E_0/\omega.
\end{align}
The Hamiltonian in Eq. \eqref{eq:two_level} is of the generic form describing strongly, periodically driven two-level system: the transverse momentum $p_y$ plays the role of a coupling strength between the two basis states \cite{PhysRevA.75.063414, Shevchenko20101}.
Other quantum systems described by this Hamiltonian include atoms in intense laser fields \cite{Shevchenko20101} and superconducting qubits \cite{Oliver2005, Son2009}.

The Floquet theorem can be applied to Eq. \eqref{eq:two_level} to obtain a formally exact solution.
The Floquet state nomenclature introduced in Son \emph{et al.} reads \cite{Son2009}
\begin{equation}
\ket{\alpha n} = \ket{\alpha} \otimes \ket{n},
\end{equation}
where $\alpha$ is the system index and $n$ is the Fourier index.
Switching to Fourier space, the Floquet eigenvalue equation reads
\begin{equation}
\sum_{\beta} \sum_{m} \bra{\alpha n} H_F \ket{\beta m} \left\langle \beta m | q_{l} \right\rangle = q_{l} \left\langle \alpha n | q_l \right\rangle,
\end{equation}
where $q_{l}$ are the Floquet quasi-energies, $\ket{q_{l}}$ are the Floquet eigenvectors, and $H_F$ is the Floquet Hamiltonian whose blocks are obtained by taking the Fourier transform of Eq.  \eqref{eq:two_level}.
Once this eigenvalue problem is solved numerically, the time-averaged transition probability between the field-free eigenstates $\ket{-}$ and $\ket{+}$ can be written as a sum of $k$-photon transition probabilities \cite{Son2009}
\begin{equation}
\label{eq:probability}
\bar{P}_{\ket{-} \rightarrow \ket{+}} = \sum_k \sum_{l} \left| \left\langle +, k | q_{l} \right\rangle \left\langle q_{l} | -, 0 \right\rangle \right|^2,
\end{equation}
where
\begin{equation}
\ket{-,k} = \frac{\varepsilon_x + |\varepsilon|}{\mathcal{N}} \ket{\alpha k} + \frac{\varepsilon_y}{\mathcal{N}} \ket{\beta k}  ,
\end{equation}
\begin{equation}
\ket{+,k} = - \frac{\varepsilon_y}{\mathcal{N}} \ket{\alpha k} + \frac{\varepsilon_x + |\varepsilon|}{\mathcal{N}} \ket{\beta k}  ,
\end{equation}
\begin{equation}
|\varepsilon| := \sqrt{\varepsilon_x^2 + \varepsilon_y^2},
\end{equation}
\begin{equation}
\mathcal{N} := \sqrt{ (\varepsilon_x + |\varepsilon|)^2 + \varepsilon_y^2}.
\end{equation}
In all numerical calculations presented in this section, the Floquet Hamiltonian is truncated to 75 blocks, for a total matrix size of $302 \times 302$.
This ensures a numerically converged solution.

In the small transverse momentum limit, i.e. $\varepsilon_y^2 \ll \varepsilon_x^2, |\mathcal{A}\omega|$, the field-free eigenstates reduce to those of $\sigma_z$ and a leading order perturbation treatment applied to the Floquet Hamiltonian leads to the following analytic formula for the transition probability \cite{Oliver2005, Son2009}
\begin{equation}
\label{eq:probability2}
\bar{P}_{\ket{-} \rightarrow \ket{+}} = \sum_k \frac{1}{2}\frac{[\varepsilon_y J_k(\mathcal{A}/\omega)]^2}{[\varepsilon_y J_k(\mathcal{A}/\omega)]^2 + [k \omega - \varepsilon_x]^2 },
\end{equation}
where $J_k$ is the Bessel function of the first kind.
In other terms, the time-averaged transition probability can be expressed as the superposition of Lorentzian $k$-photon resonances in the small transverse momentum limit.
This result can also be obtained using the adiabatic impulse model in the fast-passage limit, that is $|\mathcal{A} \omega| \gg \varepsilon_y^2$ \cite{Shevchenko20101}.

The time-dependent electron momentum density after several periods of the applied field and the time-averaged transition probability (Eq. \ref{eq:probability}) are in good agreement (see Fig. \ref{fig:density_long}).
The appearance of multiphoton rings can be seen on both results, and a similar number of cusps is obtained in individual rings with both approaches.
Consistent with Floquet theory, the low-order multiphoton resonances are broadened as the transverse momentum increases and they interact with each other, forming an intricate structure in momentum space.
As described by Son \emph{et al.}, the non-monotonical variation of the resonances' width can be directly related to the photo-induced gap between Floquet quasienergies \cite{Son2009}.
The overall time-dependent momentum pattern is also symmetrical with respect to $p_x = eE_0/\omega \simeq 47.7$ eV (or $\varepsilon_x = 0$), which is a property of the two-level Hamiltonian in Eq. \eqref{eq:two_level}.

The time-dependent and Floquet approach however differ if one considers the fast momentum space oscillations in the time-dependent momentum map (see Fig. \ref{fig:density_long} (a)).
These fast oscillations can be explained by the fact that the system only passes through a finite number of Landau-Zener transitions in the time-dependent picture, whereas in the Floquet approach the field is assumed to be periodic and applied for an infinite time.
The exact resonance condition for the fast momentum space oscillations (which are averaged out in the Floquet picture) can not, in general, be determined analytically \cite{Shevchenko20101}.
However, the oscillations are faster for a greater number of cycles (compare for instance Figs. \ref{fig:adia_vs_full_one} and Figs. \ref{fig:density_long} (a)).
They are also faster for small values of $p_y$, since the Stokes phase associated to every Landau-Zener transition, Eq. \eqref{eq:stokes_ph}, is accordingly smaller.
This smaller Stokes phase implies that the resonant values of $p_x$ are more closely spaced with decreasing $p_y$.

\begin{figure}
	\includegraphics[width=0.5\textwidth]{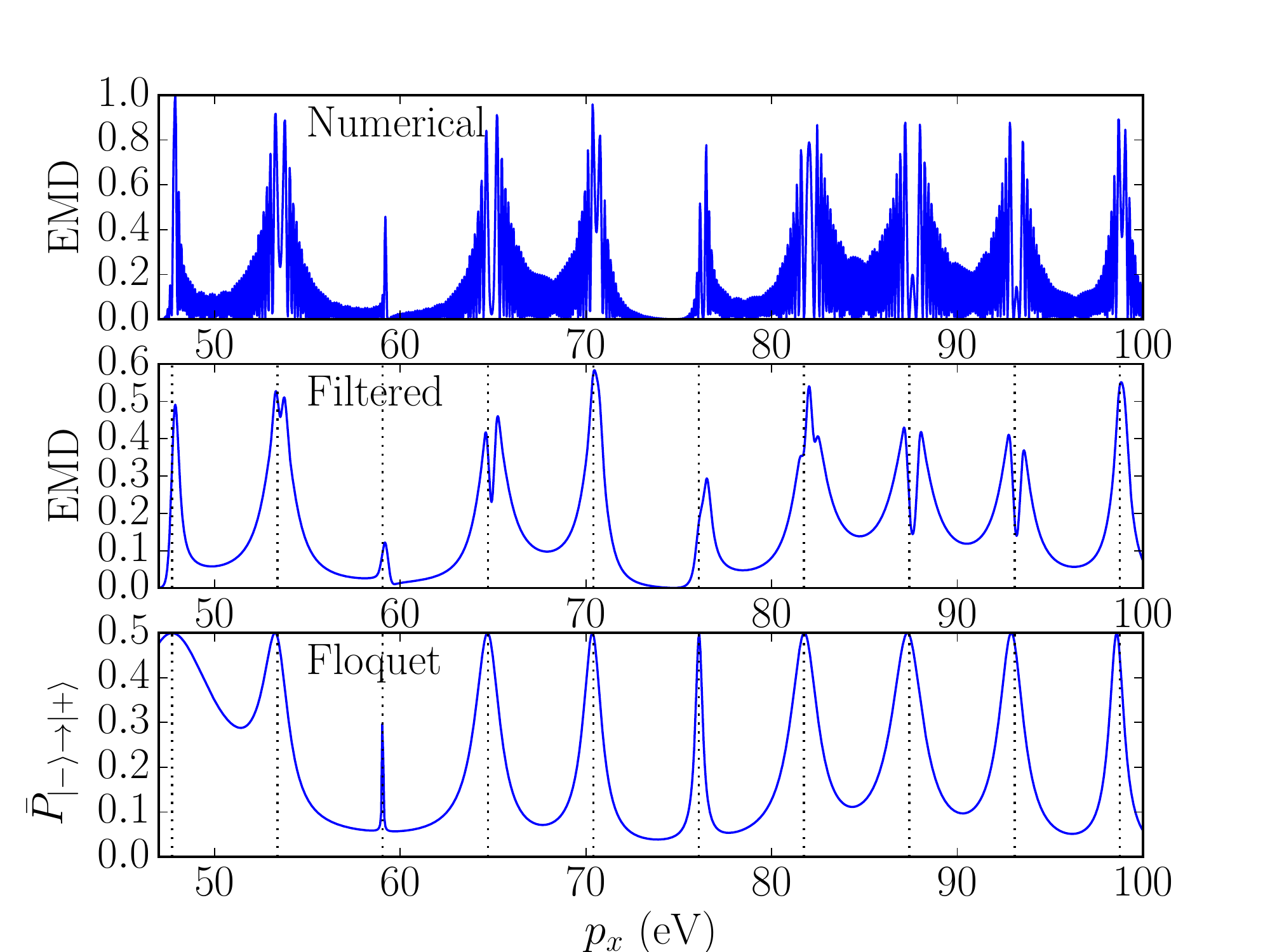}
	\caption{Comparison of the electron momentum density (EMD) per spin per Dirac point calculated via the full numerical approach after 50 cycles (top) and the transition probability computed via Floquet theory (bottom), with $p_y = 3.0$ eV.
		The electric field considered is linearly polarized in the $x$-coordinate and has a field strength of $E_{0} = 1.0 \times 10^{7}$ V/m, a frequency of $\nu = 10.0$ THz.
		A Gaussian filter with a standard deviation of 0.15 eV may be applied to the EMD to facilitate the comparison (middle).
		Dashed lines indicate the expected location of multiphoton resonances, $p_x - E_0/\omega = n \omega / 2 v_F$. }
	\label{fig:comparison}
\end{figure}

\begin{figure}
	\includegraphics[width=0.5\textwidth]{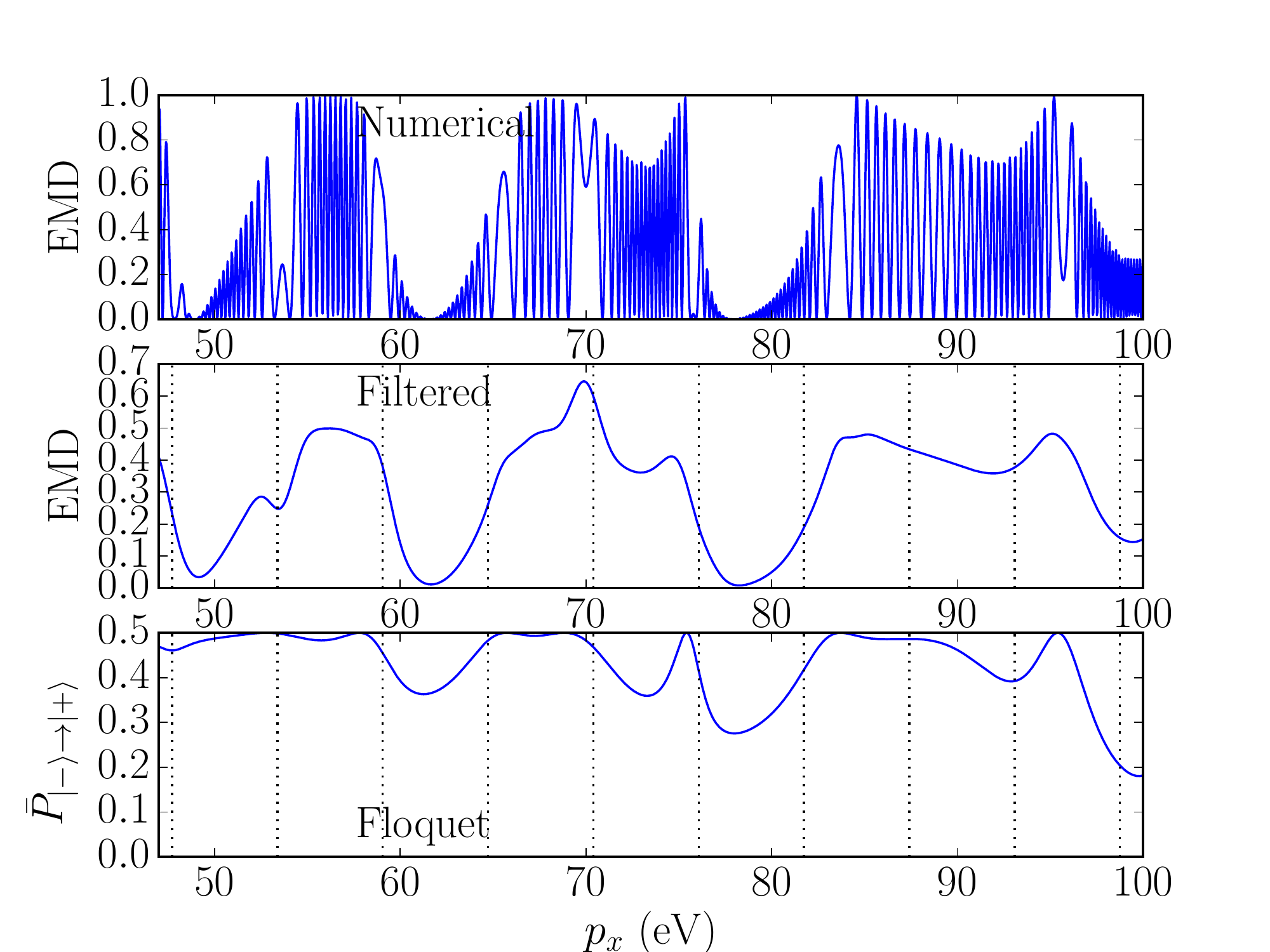}
	\caption{Comparison of the electron momentum density (EMD) per spin per Dirac point calculated via the full numerical approach after 50 cycles (top) and the transition probability computed via Floquet theory (bottom), with $p_y = 12.0$ eV.
		The applied field parameters are the same as in Fig. \ref{fig:comparison}.
		A Gaussian filter with a standard deviation of 0.6 eV may be applied to the EMD to facilitate the comparison (middle).
		Dashed lines indicate the expected location of multiphoton resonances, $p_x - E_0/\omega = n \omega / 2 v_F$. }
	\label{fig:comparison_highp}
\end{figure}


A more thorough comparison between time-dependent results and Floquet theory can be made for a fixed value of $p_y = 3.0$ eV (Fig. \ref{fig:comparison}).
The agreement is more apparent when filtering out the fast momentum space oscillations of the electron momentum density (Fig. \ref{fig:comparison}, middle panel).
Both the time-dependent electron momentum density and the time-independent result exhibit multiphoton peaks located at $p_x - eE_0/\omega \simeq n \omega / 2 v_F$ with $n$ an integer, consistent with the small $p_y$ result (Eq. \eqref{eq:probability2}).
The width of the multiphoton peaks is also well reproduced by the Floquet treatment.

The time-dependent and Floquet approach can be further compared for larger values of the transverse momentum $p_y$, but in this case the agreement is less good (see Fig. \ref{fig:comparison_highp} for $p_y = 12.0$ eV).
The pair production peaks can no longer be explained in term of a superposition of Lorentzian shaped resonances, since the condition $|\mathcal{A} \omega| \gg \varepsilon_y^2$ is no longer satisfied.
However, the shifts of the time-dependent peaks as the transverse momentum increases are qualitatively predicted by the numerical Floquet treatment.

The resonance shift at larger transverse momenta corresponds to the ac Stark effect in atomic and molecular physics.
As described by Son \emph{et al.}, the magnitude of the ac Stark effect increases with the ratio $\varepsilon_y/\varepsilon_x$  \cite{Son2009}.
In other words, lower order multiphoton ``rings'' experience a stronger Stark shift, as can be seen in Fig. \ref{fig:density_long}.
This explains the elliptic shape of the lower order resonance patterns, whereas higher order rings $(k \geq 10)$ are more circular in shape, consistent with previously obtained results for stronger fields \cite{PhysRevB.92.035401}.

\subsection{Relation to second order phase transitions}
\label{sec:second_order}

The Kibble-Zurek mechanism (KZM) gives a qualitative explanation of non-equilibrium processes occurring in second order phase transitions induced by a linear quench and has applications in cosmological and condensed matter systems \cite{0305-4470-9-8-029,KIBBLE1980183,zurek1985cosmological,Zurek1996177,doi:10.1142/S0217751X1430018X}. In particular, it predicts the density of topological defects formation after the phase transition has taken place. In this setting, the physical system is initially, at $t\rightarrow -\infty$, in a high-symmetric phase. Then, the quench drives the system across the critical point at $t=0$ and continues to $t\rightarrow \infty$. Far from the critical point, the equilibrium relaxation time $\tau$, which characterizes how fast a system returns to equilibrium when thermodynamic conditions are modified, is short, leading to adiabatic dynamics. Close to the critical point however, $\tau$ diverges: the equilibrium time is then much longer than other characteristic time scales and thus, the system is frozen. This approximate description, where adiabatic evolution is followed by frozen dynamics and adiabatic evolution again, is reminiscent of Landau-Zener transitions in two-level systems discussed in Sec. \ref{sec:adia_model}. This analogy was first noted by Damski \cite{PhysRevLett.95.035701,PhysRevA.73.063405} and was put on firm basis for the Ising model \cite{PhysRevLett.95.245701}. This was used to simulate the KZM using an optical interferometer \cite{PhysRevLett.112.035701} and superconducting qubits \cite{gong2016simulating}. In the adiabatic limit considered in this article, graphene can also be used as a ``non-equilibrium physics simulator'' owing to the description of quasi-particles in terms of Landau-Zener transitions.

This connection can be made explicit by following the discussion given in Ref. \cite{PhysRevLett.95.035701}. First, the relaxation time is related to the inverse of the gap as  
\begin{eqnarray}
\label{eq:relax_time}
\tau(t) := \frac{1}{\Delta(t)} = \frac{1}{v_{F}\sqrt{[eE_{x}(t_{j})t]^{2}+ p_{y}^{2}}},
\end{eqnarray}
close to the $j$th nonadiabatic transition. Using this definition in the quantum setting for graphene, the relaxation time is large in the vicinity of nonadiabatic transitions where the system is effectively frozen, analogously to the thermodynamic setting. Zurek's equation then reads \cite{PhysRevLett.95.035701}
\begin{eqnarray}
\label{eq:zurek}
\tau(\hat{t}) = \xi \hat{t},
\end{eqnarray} 
where $\xi = \pi/2$ \cite{PhysRevA.73.063405} and $\hat{t}$ is the freeze-out time that determines when the system switches from an adiabatic to a nonadiabatic evolution. In other words, for $t \in [t_{j}-\hat{t},t_{j}+\hat{t}]$ for $j =1,\cdots,n$, the dynamics is nonadiabatic while for every other times, it is adiabatic. 

\begin{figure}
	\includegraphics[width=0.5\textwidth]{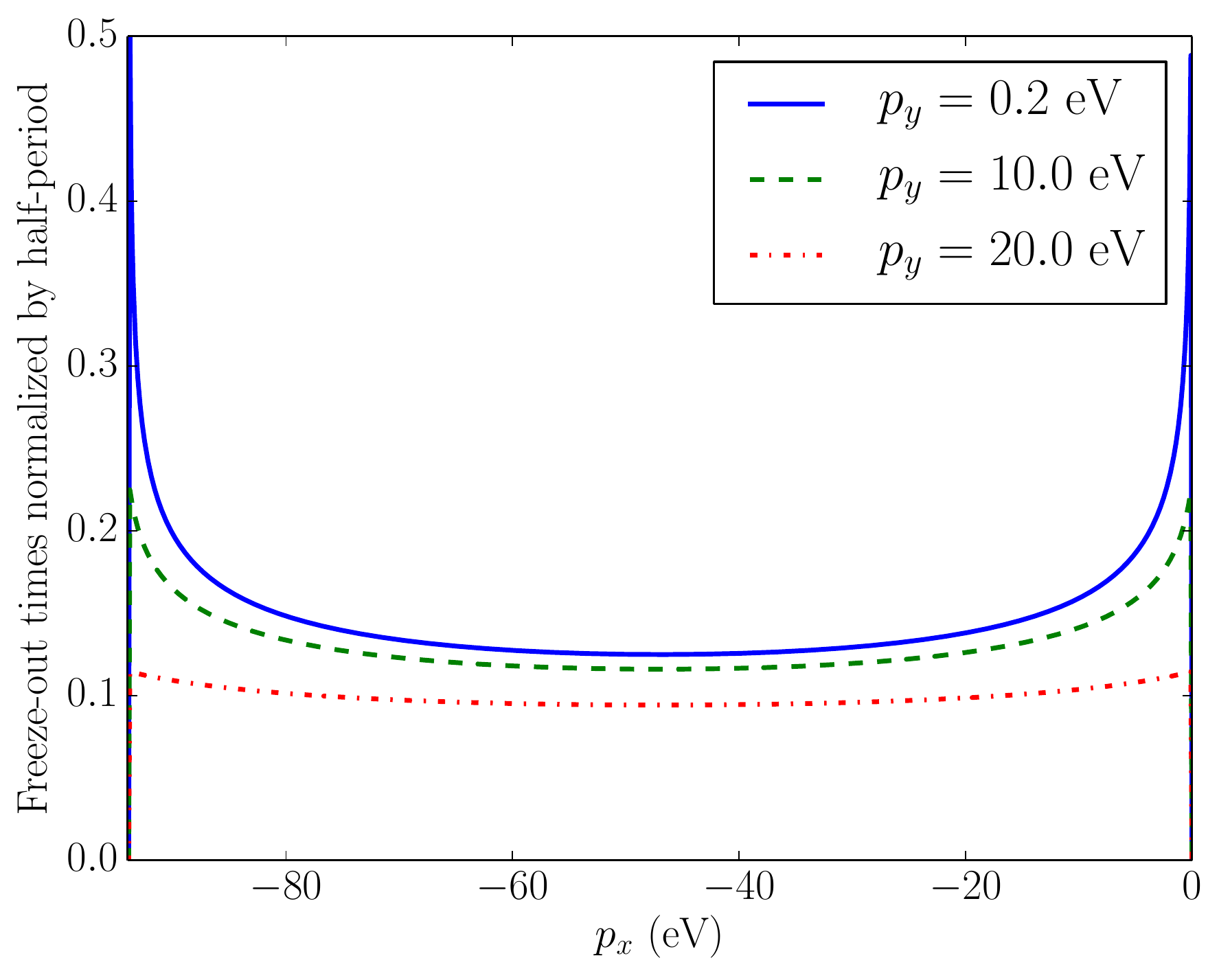}
	\caption{Numerical results for the normalized freeze-out time $\hat{t}/T_{1/2}$, where $T_{1/2}$ is a half-period of an oscillating external field linearly polarized in the $x$-coordinate, with a field strength of $E_{0} = 1.0 \times 10^{7}$ V/m and a frequency of $\nu = 10.0$ THz.}
	\label{fig:freeze_out}
\end{figure}

A solution to Eq. \eqref{eq:zurek} can be found and is given by
\begin{eqnarray}
\hat{t}_{j} = \frac{\tau_{Q}^{(j)}}{\sqrt{2}} \sqrt{\sqrt{1+ \frac{4\tau_{0}^{2}}{\xi^{2}(\tau_{Q}^{(j)})^{2}}}-1},
\end{eqnarray} 
where
\begin{eqnarray}
\label{eq:quench_time}
\tau_{0} := \frac{1}{v_{F}|p_{y}|}, \quad
\tau_{Q}^{(j)} := \frac{|p_{y}|}{eE_{x}(t_{j})}.
\end{eqnarray}
Here, $\tau_{0}$ is a constant that characterizes the relaxation time and $\tau_{Q}^{(j)}$ is the quench time scale. The freeze-out time can be computed for the electric field considered in Sec. \ref{sec:quant_inter} (a), for one half-cycle. The numerical result is displayed in Fig. \ref{fig:freeze_out} where the normalized freeze-out time is given for all momenta $p_{x}$ considered for the adiabatic-impulse model calculations in Sec. \ref{sec:quant_inter}. Close to $p_{x} \approx 0.0$ and $p_{x} \approx -95.4$ eV, the freeze-out time becomes large, of the same order as the half-period. This also explains the discrepancy between exact numerical results and the ones obtained from the adiabatic-impulse model: close to $p_{x} \approx 0.0$ and $p_{x} \approx -95.4$ eV, the freeze-out or nonadiabatic behavior lasts for almost all the half-cycle. Therefore, the dynamics does not proceed by a sequence of adiabatic evolution and nonadiabatic transitions, as assumed in the adiabatic-impulse model. Rather, it is always in the nonadiabatic regime, resulting in less accurate results.  

Once the time scales relating thermodynamic systems to graphene are defined, as given in Eqs. \eqref{eq:relax_time} - \eqref{eq:quench_time}, it is possible to interpret the electron momentum density as the density of topological defects. Using scaling laws, it has been demonstrated that the density of topological defects for a quenched quantum Ising model scales like \cite{PhysRevLett.95.105701}
\begin{eqnarray}
\label{eq:topo}
\langle n_{\mathrm{topo}}\rangle \sim \sqrt{\frac{\tau_{0}}{\tau_{Q}}},
\end{eqnarray}
where here, $\tau_{Q}$ is the quench time of the Ising model, analogously to Eq. \eqref{eq:quench_time}.
In graphene, the same scaling can be found by looking at the electron momentum density obtained after one nonadiabatic transition, given in Eq. \eqref{eq:pair_prod_adia_one}. Integrating the latter on the transverse momentum $p_{y}$ and assuming $p_{x} \in [-eA_{x,\mathrm{in}},0]$, we get
\begin{eqnarray}
\left. \frac{d\langle \tilde{n}_{s,a} \rangle}{dp_{x}}\right|_{p_{x}\in [-eA_{x,\mathrm{in}},0]} = \sqrt{\frac{eE_{x}(t_{j})}{v_{F}p_{y}^{2}}} = \sqrt{\frac{\tau_{0}}{\tau_{Q}^{(j)}}}.
\end{eqnarray}
This result for the electron momentum density at a given $p_{x}$ is consistent with the scaling of defects in the Ising model, Eq. \eqref{eq:topo}, confirming the analogy between the adiabatic dynamics of quasiparticles in graphene and topological defect production in second order phase transitions. A similar result was found in Ref. \cite{PhysRevB.81.165431}.

\section{Conclusion}
\label{sec:conclu}

In this article, the electron momentum density in graphene created by an external classical electric field was computed using numerical methods combined with techniques borrowed from strong field QED. Several time dependences of the applied field have been studied in the tunneling/Schwinger regime where $\gamma \ll 1$. It was demonstrated that when the system is driven periodically, nonadiabatic transitions occur when the adiabatic mass gap is minimal, resulting in a quantum interference pattern in the pair momentum density, reminiscent of Landau-Zener-St\"{u}ckelberg interferometry. This interpretation was confirmed by using the adiabatic-impulse model, which corrects the full adiabatic evolution by adding nonadiabatic transitions when the adiabatic mass gap is minimal.

In this adiabatic limit, the production of electron-hole pairs in graphene is analogous to the generation of topological defects in quenched second order transitions. Using estimates obtained from the analysis of the quenched quantum Ising model, it was possible to evaluate the graphene analog to the freeze-out time. Using this freeze-out time, it was possible to explain the discrepancy between exact numerical results and the ones obtained from the adiabatic-impulse model: in some momentum regions, $\hat{t}$ is of the same order of magnitude as the half-period, meaning that the system is never adiabatic, contrary to the assumption in the adiabatic-impulse model. Finally, comparing again to results obtained for the Ising model, it was demonstrated that the defect density is analogous to the electron momentum density in graphene at fixed $p_{x}$. Therefore, Schwinger-like pair production in graphene could be used as a simulator for the Kibble-Zurek mechanism, in the same spirit as some recent experimental investigations using superconducting qubit systems \cite{gong2016simulating}.

In the long time limit, the system goes through many avoided crossings. As a consequence, an intricate interference pattern appears in the pair momentum density. Its time-averaged features can be understood in the low transverse momentum limit by introducing the Floquet formalism. In particular, in the limit of a large number of cycles, the electron momentum density exhibits multiphoton rings which are formed by the sequential nonadiabatic transitions. Clearly, the rings appear for momenta where constructive interference occurs.
Destructive interference, on the other hand, is associated with the phenomenon of coherent destruction of tunneling \cite{Grifoni1998229, Shevchenko20101}.

It is interesting to see the appearance of multiphoton rings in the long time limit as these are usually understood as a signature of the multiphoton regime where $\gamma \gg 1$ \cite{PhysRevB.92.035401}. Our study shows that multiphoton rings are also present in the tunnelling regime as a result of quantum interference. Therefore, it is possible that $\gamma$ characterizes how rapidly multiphoton rings come into existence. This will be investigated further in other studies.

\appendix

\section{Crossing of resonances and minimal gap}
\label{app:min_gap}

The complex times when there is a crossing of the adiabatic energies are solutions of
\begin{eqnarray}
E_{\mathbf{p}}^{2}(\Omega t^{*}) = v_{F}^{2}\left[ p_{x} + eA_{0} + eA(\Omega t^{*}) \right] + v_{F}^{2}p_{y}^{2} = 0,
\end{eqnarray}
where $\Omega$ is the typical time scale of the vector potential. Of course, because the time $t^{*}$ is complex, the vector potential is a complex-valued function $A(t^{*}) \in \mathbb{C}$. Splitting the real and imaginary parts as $A(t^{*}) = A_{\mathrm{R}}(t^{*}) + i A_{\mathrm{I}}(t^{*})$, we obtain two equations:
\begin{eqnarray}
\label{eq:sol_im1}
\left[ p_{x} + eA_{0} + eA_{\mathrm{R}}(\Omega t^{*}) \right]A_{\mathrm{I}}(\Omega t^{*}) &=& 0 ,\\
\label{eq:sol_im2}
\left[ p_{x} + eA_{0} + eA_{\mathrm{R}}(\Omega t^{*}) \right]^{2} - e^{2}A_{\mathrm{I}}^{2}(\Omega t^{*}) + p_{y}^{2} &=& 0.
\end{eqnarray}
For the following, it is convenient to express the complex time as $t^{*} = t_{\mathrm{R}} + it_{\mathrm{I}}$. Then, as demonstrated in Eq. \eqref{eq:cond1}, the minimal gap occurs when the condition
\begin{eqnarray}
\label{eq:min_gap}
p_{x} + eA_{0} + eA_{\mathrm{R}}(\Omega t_{\mathrm{R}}) = 0,
\end{eqnarray}
is fulfilled. Here, we assume that the imaginary part is small such that $\Omega t_{\mathrm{I}} \ll 1$, allowing for an expansion of the vector potential as
\begin{eqnarray}
A_{\mathrm{I}}(\Omega t^{*}) &=& A_{\mathrm{I}}(\Omega t_{\mathrm{R}}) +   t_{\mathrm{I}} \left[\partial_{t_{\mathrm{I}}} A_{\mathrm{I}}(\Omega t^{*})\right]_{t_{\mathrm{I}}=0} \nonumber \\
&& + O(\Omega^{2} t_{\mathrm{I}}^{2}), \\
&=& t_{\mathrm{I}} \left[ \partial_{t_{\mathrm{R}}} A_{\mathrm{R}}( t^{*})\right]_{t_{\mathrm{I}}=0}, \\
\label{eq:AI}
&=& -  t_{\mathrm{I}} E_{x}(t_{\mathrm{R}}),
\end{eqnarray} 
where the second equation is obtained from the Cauchy-Riemann equations and from the fact that $A_{\mathrm{I}}(\Omega t_{\mathrm{R}}) = 0$. A similar argument can be performed for $A_{\mathrm{R}}$ and it can be shown that
\begin{eqnarray}
\label{eq:AR}
A_{\mathrm{R}}(\Omega t^{*}) &=& A_{\mathrm{R}}(\Omega t_{\mathrm{R}}) +  O(\Omega^{2} t_{\mathrm{I}}^{2}),
\end{eqnarray}
using the fact that $\mathrm{Im}\left[E_{x}(\Omega t_{\mathrm{R}}) \right] = 0$. Reporting the result of Eqs. \eqref{eq:AI} and \eqref{eq:AR} into Eqs. \eqref{eq:sol_im1} and \eqref{eq:sol_im2}, along with the minimal gap condition (Eq. \eqref{eq:min_gap}), we get the solution
\begin{eqnarray}
t_{\mathrm{I}} = \frac{|p_{y}|}{e|E_{x}(\Omega t_{\mathrm{R}})|} + O(\Omega^{2}t_{\mathrm{I}}^{2}).
\end{eqnarray}
Then, the condition to expand the vector potential becomes
\begin{eqnarray}
\Omega t_{\mathrm{I}} = \frac{\Omega |p_{y}|}{e|E_{x}(\Omega t_{\mathrm{R}})|}  = \gamma \ll 1,
\end{eqnarray}
consistent with the tunnelling regime given in Eq. \eqref{eq:keldysh}. Finally, $t_{\mathrm{I}}$ is related to the transition probability \cite{Shevchenko20101} and it can be shown that it yields the parameter $\delta_{j}$ defined in Eq. \eqref{eq:delta}. As a consequence, non-adiabatic transitions really occur when the gap is minimal, up to corrections $O(\gamma^{2})$. This also confirms the validity of the adiabatic-impulse approach in this regime. However, there are some parameters where the reasoning presented in this Appendix does not hold, in particular when the gap is not minimized by the condition Eq. \eqref{eq:min_gap} but rather, by $E_{x}(t)=0$ (see Eq. \eqref{eq:cond2}). In this case, the crossing of adiabatic energies and the minimum gap occur at different times. Then, other more sophisticated techniques have to be employed \cite{PhysRevLett.104.250402,PhysRevD.83.065028,PhysRevA.44.4280}.

\begin{acknowledgments}
The authors are thankful to J. Dumont and P. Blain for useful discussion and comments. One of the authors (FFG) is grateful to S. Deffner for pointing out the existence of the Kibble-Zurek mechanism.
Computations were made on the supercomputer MAMMOUTH from Universit\'{e} de Sherbrooke, managed by Calcul Qu\'{e}bec and Compute Canada.
The operation of this supercomputer is funded by the Canada Foundation for Innovation (CFI), the minist\`{e}re de l'\'{E}conomie, de la science et de l'innovation du Qu\'{e}bec (MESI) and the Fonds de recherche du Qu\'{e}bec -- Nature et technologies (FRQNT).
\end{acknowledgments}

\bibliographystyle{apsrev}

\bibliography{bibliography}

\end{document}